\documentclass{aa}

\usepackage{graphicx}
\usepackage{txfonts}

\usepackage[colorlinks=true,urlcolor=blue,citecolor=red,linkcolor=red,bookmarks=true]{hyperref}

\usepackage{amsmath}
\usepackage{gensymb}

\def\e#1{{\em #1}}

\begin{document}

\title{The young Hobson family: \\ Possible binary parent body and low-velocity
       dispersal} 

\author{D. Vokrouhlick\'y\inst{1},
        M. Bro\v{z}\inst{1},
        B. Novakovi\'{c}\inst{2},
        and D. Nesvorn\'y\inst{3}}

\titlerunning{The young Hobson family}
\authorrunning{Vokrouhlick\'y et~al.}

\institute{Institute of Astronomy, Charles University, V~Hole\v{s}ovi\v{c}k\'ach 2,
           CZ-180~00 Prague 8, Czech Republic \\ \email{vokrouhl@cesnet.cz,
           mira@sirrah.troja.mff.cuni.cz}
      \and
           Department of Astronomy, Faculty of Mathematics, University of Belgrade,
           Studentski trg 16, 11000 Belgrade, Serbia 
      \and
           Southwest Research Institute, 1050 Walnut St, Suite 300,
           Boulder, CO 80302, USA}

\date{Received: \today ; accepted: ???}

\abstract
{Asteroid families with ages younger than $1$~Myr offer an interesting possibility
 of studying the outcomes of asteroid disruptions that are little modified by
 subsequent evolutionary processes.}
{We analyze a very young asteroid family associated with (18777) Hobson in the central part of
 the main belt. We aim at (i) understanding its peculiar size distribution, and (ii)
 setting an upper limit on the characteristic dispersal velocity at subkilometer sizes
 corresponding to the smallest visible Hobson members.}
{We identified the Hobson family using an up-to-date asteroid catalog. A significant increase
 in the number of its known members allowed us to study their size distribution and compare it with
 computer simulations of catastrophic disruptions. Backward orbital integrations of the
 heliocentric orbits allowed us to confirm the previously suggested age of Hobson and helped to
 estimate limits of the ejection speed.}
{The Hobson family has an unusual size distribution: two nearly equal-size bodies, followed by
a population of smaller asteroids, whose distribution takes a characteristic power law.
 There are two possibilities to explain these data. Either a canonical impact onto a single parent body,
 requiring fine-tuned impact conditions that have not been studied so far, or an unconventional model
 for the parent object of the Hobson family, namely a binary with $\simeq 7-9$~km primary and a
 $\simeq 2.5$~km secondary. In the latter case, the primary was disrupted, leaving behind the largest
 remnant (18777) Hobson and a suite of subkilometer asteroids.
 The second largest asteroid, (57738) 2001~UZ160, is the nearly intact satellite of the parent
 binary. The excellent convergence of nominal orbits of Hobson members sets an upper limit of 
 $\simeq (10-20)$ m~s$^{-1}$ for the initial dispersal velocity of the known members, which is
 consistent with both formation models. The Hobson family provides a so far rare opportunity
 of studying disruptions of small asteroids in a situation in which both the material strength and
 reaccumulation efficiency play an important role.}
{}


\maketitle


\section{Introduction} \label{intr}
Analyses of asteroid families, which consist of the dispersed pieces of a parent body that experienced a
strong impact, belong to the fundamental tools that allow us to study asteroids in planetary
science. By telescopic observations of the identified fragments, the families offer
a look into a previously existing body and thereby allow us to constrain the degree of its thermal
processing and geophysical differentiation. Studying the configuration of asteroids
constituting a family, researchers may constrain the fragmentation process of the parent 
body. Statistics of families with parent bodies of different sizes in a given
population (such as the main belt) may help constrain the general lines of its
collisional evolution. The dynamical processes that may affect the configuration of
fragments in a given family after it formed may provide information about its age.
Some families may be suitably situated to instantly deliver large number of fragments
to some of the principal resonant escape hatches to the planet-crossing population
and therefore potentially affect the impactor flux onto terrestrial planets in
the past. Kiyotsugu Hirayama would likely be surprised by the immense importance%
\footnote{A conclusion against some early skeptical opinions, such as expressed by the
 late Ernest Brown: {\em \textup{\textup{``Any hope of getting evidence from this source
 as to whether they [i.e. Hirayama families] were a result of an explosion
 or collision must be abandoned.'}'} \citep{b1932}} }
of the asteroid families, about a century after he discovered the first three
examples \citep{hira1918}.

The parameters of the size distribution and the velocity field with which the fragments
were dispersed at the family-formation event are one of the data that researchers
are trying to determine from the configuration of the fragments. This provides interesting information
not only about the mechanics of the parent-body fragmentation, but may potentially provide
important clues for the efficiency with which specific families were capable of directly
delivering sizable fragments into resonances and thus set them on track to the
planet-crossing populations. Hirayama himself considered this possibility \citep{hira1928},
see also \citet{b1950,b1951}, and obtained dispersal velocities of several hundreds of
meters per second. These early works were able to grasp just the core of the families,
consisting of their largest members, which means that the dispersal velocities of smaller members
would likely be even higher. Modern versions of these efforts could have started only
after (i) a much larger number of asteroids had been discovered, and (ii) tools for
determining accurate values of the proper orbital elements were developed and became
widely accessible. These two criteria are met since the early 1990s,
and \citet{zetal1996} and \citet{cetal1999} were among the first to apply techniques
for estimating the dispersal velocity field from the observed families. However,
these works only confirmed the earlier-noted problem that the determined velocities were systematically too
high \citep[e.g., ~ ][]{zetal2002}.
At the same time, a solution of the discrepancy was at hand and consisted of the
recognition of the dynamical processes that in the long term disturb the proper
orbital elements that are used in the methods for determining velocity fields methods mentioned above.
These processes primarily are of two types: (i) size-independent chaotic dynamics in diffusive mean-motion
resonances \citep[e.g.,][]{flora2002}, and (ii) size-dependent dispersion due to the
Yarkovsky effect \citep[e.g.,][]{fv1999,betal2001}. In brief, the configuration of many
asteroid families in the space of proper orbital elements is too inflated by the
dynamical perturbations, such that the information about a much smaller initial state that is
directly related to the dispersal velocity field is basically effaced. While some
attempts to remove the dynamical component exist
\citep[e.g.,][]{vetal2006}, their accuracy becomes quite low as soon as the
family age exceeds several tens or hundreds of million years.

The conditions required to overcome the dynamical evolution problem is apparent: the
time to perturb the family configuration must have been short, or in other words, young
families must be studied. While evident, the plan has its own difficulties. First, recent breakups
would statistically imply smaller parent bodies, and thus also smaller sizes of
the currently observed fragments. This again creates the need to extent reliable information
about asteroid populations to smaller objects. Second, a chosen family needs to be confirmed 
to be young. An important step toward
young asteroid family science occurred in 2002 with discovery of the Karin family
\citep{karin2002}. In addition to an increased number of known asteroids in the available
catalogs, the key novelty of this work consisted of the idea that the origin of the
family was dated by convergence of the secular angles, longitude of node $\Omega$ and perihelion
$\varpi$, of the identified members. The families are by definition tight clusters
in semimajor axis, eccentricity, and inclination values. This is how they
are identified in the first place. However, the values of secular angles are
generally randomized by differential precession. At their origin, even the
$\Omega$ and $\varpi$ values must have been clustered, however. Therefore, ideally,
past simultaneous convergence of the secular angles of a significant number of
members unambiguously signals the epoch of the family origin. \citet{karin2002}
studied  $13$ suitable Karin members and reported that this family is $5.8\pm 0.2$~Myr old
\citep[the result may even be improved using the increasing number of recognized
Karin members; e.g.,][]{karin2004,karin2016}. More details about the Karin family,
especially about what it can tell us about the initial dispersal velocities of its
members, are given in Sec.~\ref{disc}.

Karin has started a new era of intense search for young asteroid families.
Experience showed that the secular-angle convergence technique can be applied
for families whose age is younger than $\simeq (10-15)$~Myr \citep[several
examples are given in, e.g.,][]{netal2015}.
A new twist on the way to this goal was reached by the discovery of very young clusters with
ages younger than $\simeq 1$~Myr. These extremely young structures possess an additional
quality compared to the older families: their osculating (or mean) secular
angles at the current epoch are still clustered because their differential precession
did not have enough time to disperse them since the family origin. This
property plays an important role in their identification (see also Sec.~\ref{iden}).
These very young clusters therefore represent an even more pristine state that
potentially allows us to characterize details of their formation event, including the
velocity dispersal parameters. The first examples were discovered by \citet{datura2006}
and \citet{nv2006}. While fascinating, many of the very young clusters are still 
poorly characterized because only few of their members are discovered so far. For only three 
cases are more than $\sim 20$ members known well enough to allow serious analysis:%
\footnote{Current but unpublished counts of the membership in the Datura and Schulhof
 families indicate $63$ and $27$ fragments, while the recent census of the Adelaide
 family has $59$ fragments.} 
(i) The Datura family \citep[e.g.,][]{datura2009,datura2017}, (ii) the Schulhof
family \citep[e.g.,][]{vetal2011,vetal2016}, and (iii) the Adelaide family
\citep[e.g.,][]{vetal2021}.
The Datura and Adelaide families are examples of large cratering events, such that
the largest observed fragment in the family gives us a good idea about the parent
body. The Schulhof family is at the border between the cratering and catastrophic
fragmentation of the parent object. None of the three cases of very young clusters
with a wealth of fragments is the result of a highly energetic collision between
a projectile and a parent body. This is expected because an impact by a small projectile 
is more likely than an impact by a large projectile, as is required for a super-catastrophic
collision. This situation would be perhaps be the most interesting as far as
the information about the velocity dispersal field is concerned, however. As
a rule of thumb, the cratering events lift the escaping fragments with a characteristic
velocity similar to the escape velocity from the parent object.

With this motivation in mind, we analyze the Hobson family.
A tight cluster of four asteroids about the middle main belt object, (18777) Hobson
was first reported by \citet{pv2009} as a side product of their search for
asteroid pairs. These authors therefore did not pay much attention to
its analysis other then noting the young age, younger than $0.5$~Myr, based on the mutual
convergence of their orbital secular angles. Rosaev and Pl{\'a}valov{\'a} then
studied the Hobson cluster in a series of papers \citep[see][]{rp2016,rp2017,rp2018}
with the following principal conclusions: (i) with more complete datasets of
asteroids available, they completed the membership in this family up to nine
members by 2018, (ii) they recognized the chaotic nature of the orbital evolution
for most of the family members and pointed out a possible role of perturbations
by the dwarf planet Ceres, and (iii) they used a simple model to constrain the Hobson family
age, obtaining $365\pm 67$~kyr. Finally, the Hobson cluster was briefly analyzed by
\citet{petal2018},
who added two more members (pushing the count to 11 asteroids) and
confirmed the age to between $300$ and $400$~kyr with a different approach.
They also concluded that the Hobson family is an outlier in their search for clusters
possibly formed by a rotational fission of a parent asteroid. In all 
likelihood, it must have formed in a more traditional way, notably by catastrophic
collision of two asteroids. Finally, they conducted valuable photometric observations
of the two largest members, (18777)~Hobson and (57738)~2001~UZ160 (Sec.~\ref{lfs}).
\begin{figure*}[t]
 \begin{center} 
 \includegraphics[width=0.85\textwidth]{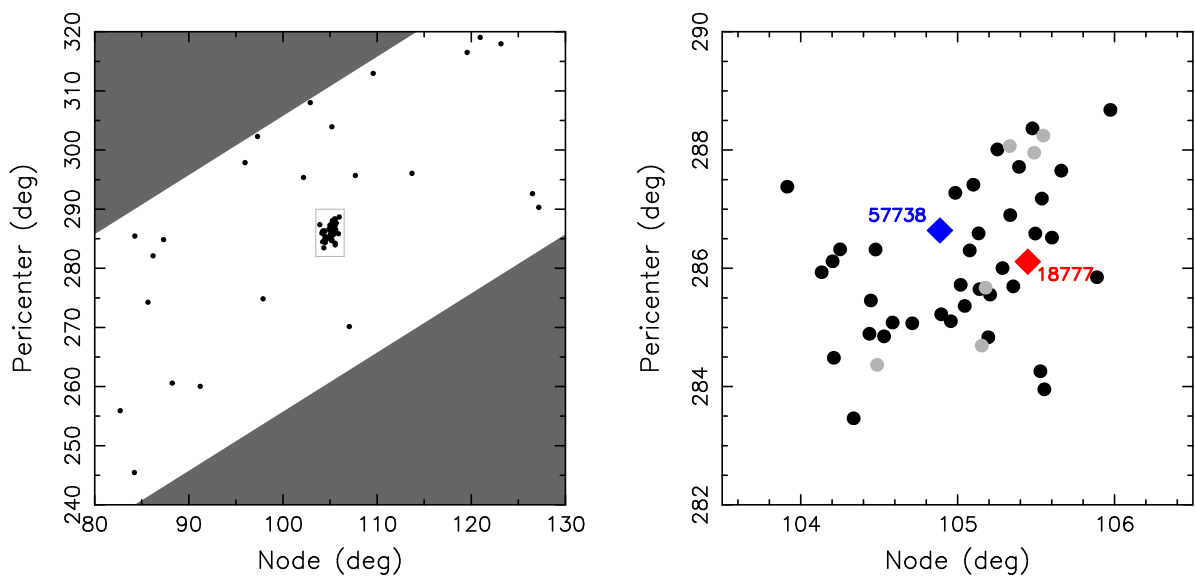}
 \end{center}
 \caption{\label{f1}
  Osculating values of the heliocentric secular angles in the ecliptic system in the target
  zone surrounding the Hobson family: the longitude of node $\Omega$ (abscissa) vs.
  the longitude of perihelion $\varpi$ (ordinate). {\em Left panel}: Total of $69$
  asteroids, clearly divided into (i) a random background population, and (ii) a strongly
  clustered family (red symbols highlighted by the rectangle). Because the initial search used the
  argument of perihelion $\omega$, the actively probed region around (18777) Hobson
  becomes tilted in $\Omega$ vs. $\varpi$ axes, and the corner triangles in dark gray are
  excluded. {\em Right panel}: Zoom on the Hobson family structure (range of the axes
  as in the gray rectangle in the left panel). The two largest asteroids, (18777) Hobson
  and (57738) 2001~UZ160, are highlighted in red and blue with diamonds. All smaller
  members are shown by filled circles: multi-opposition orbits in black, and the six
  single-opposition orbits in gray (asteroids 2014~JH120, 2014~OJ66, 2020~JM31, 2020~KP36,
  2020~OY50, and 2021~MO5; Table~\ref{tab_prop}).
  The family is unusually compact in both $\Omega$ and $\varpi$, promising very young age.}
\end{figure*}
\begin{figure}[t]
 \begin{center} 
 \includegraphics[width=0.49\textwidth]{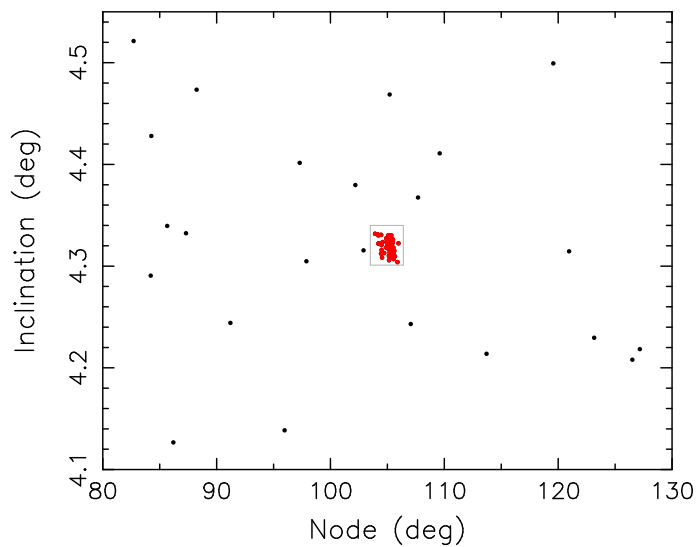}
 \end{center}
 \caption{\label{f2}
  Osculating values of the longitude of node $\Omega$ (abscissa) vs. the inclination $I$
  (ordinate) in the target zone surrounding the Hobson family. The limits of the axes
  depict our search zone. The members of the Hobson family (red symbols in the rectangle)
  are strongly clustered and can easily be distinguished from the dispersed background
  population of asteroids.}
\end{figure}

Taking up this preliminary information, we revisit the Hobson family by noting a
significant increase in the population of its fragments in the updated asteroid
catalogs. Our count indicates a total of $45$ members (Sec.~\ref{iden}). This
relatively large sample allows us to improve the analysis and argue that the Hobson family
probably is the result of a collision of a kilometer-sized projectile and a target that was about ten times
larger (Secs.~\ref{sfd} and \ref{sfd_mod}). Interestingly, the size
distribution of the Hobson fragments presents nontrivial difficulties. It either
requires a low-probability parametric combination that was not sampled in previous studies, or
even a novel idea about is parent object: a typical small binary asteroid in the inner
main belt population (Secs.~\ref{sfd} and \ref{sfd_mod}). The analysis of the past convergence
of the heliocentric orbits of the Hobson members provides information about the characteristic
dispersal velocity at the origin (Sec.~\ref{conv}), and this appears to be in accord
with both formation models. The regime of kilometer-size
parent objects colliding at high speed has not been probed yet in the available
family data (Sec.~\ref{disc}). Our census of family members, together with their
proper elements, is given in Appendix~A. Some details related to our simulations
of small-asteroid breakups are provided in Appendix~B.

\section{Hobson family} \label{fam}

\subsection{Largest fragments} \label{lfs}
The Hobson family is located in the middle part of the main belt and consists of
small asteroids. As a result, we have only very little information about their
physical properties. Fortunately, \citet{petal2018} conducted calibrated photometric
observations of the two largest members in the season 2013/2014 with the
following results:
\begin{itemize}
\item The larger fragment, (18777)~Hobson, had a synodic rotation period of $10.227\pm
 0.004$~hr and a rather low light-curve amplitude of $0.21$~magnitude. Observations
 at different filters provided a color index $V-R=0.477\pm 0.010$~magnitude (using the
 Johnson-Cousins standard system) and an absolute magnitude $H=15.16\pm 0.05$ at the middle of the light curve.
\item The smaller fragment, (57738)~2001~UZ160, had a synodic rotation period of $20.51\pm
 0.01$~hr and a larger light-curve amplitude of $0.65$~magnitude. Observations
 at different filters provided a color index $V-R=0.46\pm 0.02$~magnitude
 and an absolute magnitude $H=15.41\pm 0.05$ at the middle of the light curve.
\end{itemize}
The slow rotation periods are notable, and the information about statistically compatible
values of the color index $V-R$ is important. They are consistent with S-type classification,
which is not surprising in this zone of the asteroid main belt.
We can thus extrapolate this result and consider that the Hobson family belongs to the
S-type group. The geometric albedo value of S-type asteroids is $p_{\rm V}=0.20\pm 0.05$
\citep[e.g.,][]{petal2012}, and this helps us to estimate that the sizes of (18777)~Hobson
and (57738) 2001~UZ160 are $2.82\pm 0.39$~km and $2.52\pm 0.32$~km (considering an 
uncorrelated uncertainty in absolute magnitude and geometric albedo for simplicity).

\subsection{Family identification} \label{iden}
In order to identify members in the Hobson family in current asteroid catalogs, we used 
the straightforward approach of \citet{vetal2021}. Following
previous indications of its very young age, we assumed that the family must be clustered in
a five-dimensional space of the osculating orbital elements: semimajor axis $a$, eccentricity
$e$, inclination $I$, longitude of node $\Omega,$ and longitude of perihelion $\varpi$
(alternatively, argument of perihelion $\omega$). The higher dimensionality of
this space, over just three dimensions of the proper orbital element space in which
asteroid families are typically identified \citep[e.g.,][and Appendix~A]{netal2015}, 
helps us to unambiguously discern the family members from background population of
asteroids. We considered (18777)~Hobson as a point of reference and constructed a 
box zone around it by letting the osculating orbital
elements vary in some range: (i) semimajor axis $\pm 0.03$~au, (ii) eccentricity
$\pm 0.03$, (iii) inclination $\pm 0.2^\circ$, (iv) longitude of node $\pm 25^\circ$,
and (v) argument of perihelion $\pm 25^\circ$. The adopted range of values is
an order of magnitude (or significantly more in secular angles) larger than the 
short-period oscillations of the osculating orbital elements of asteroids in this zone.
We used the catalog of asteroid orbital
elements provided by Minor Planet Center ({\tt MPCORB.DAT}) as of July~15, 2021,
which contained about $1\, 100\, 000$ entries, and we only discarded very poorly characterized
single-opposition orbits with an observation arc shorter than a week.
\begin{figure*}[t]
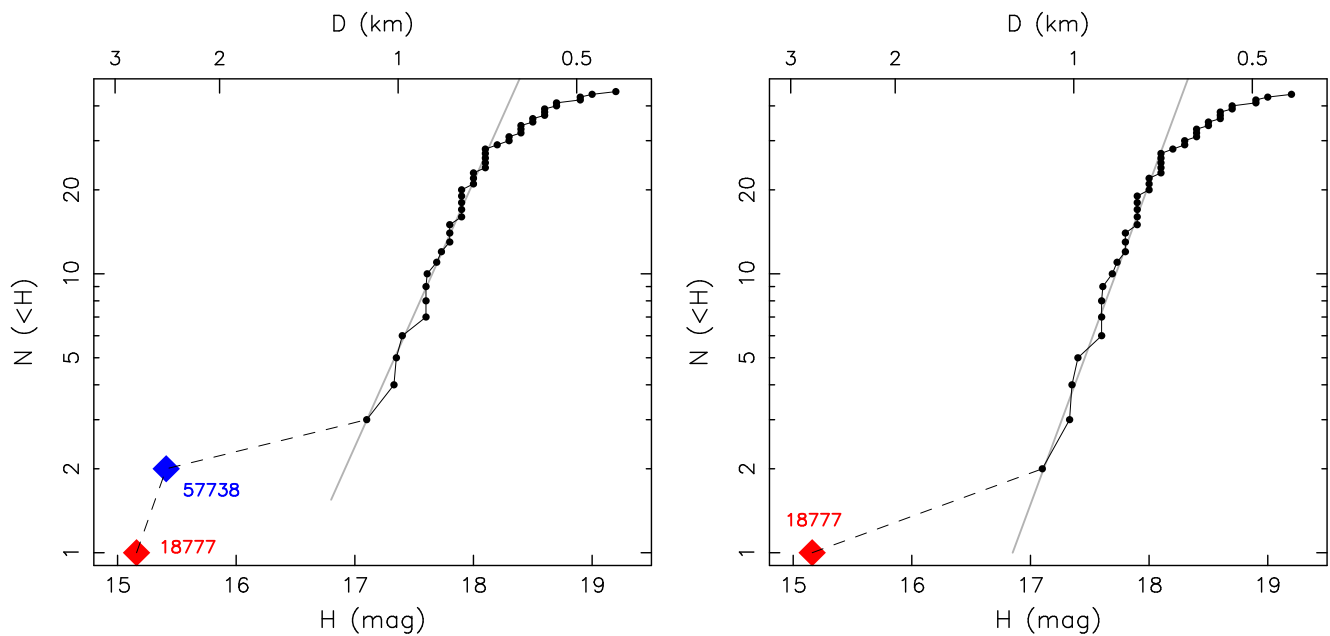

 \begin{center}
  \begin{tabular}{cc}   
   \includegraphics[width=0.46\textwidth]{f3a.eps} &
   \includegraphics[width=0.46\textwidth]{f3b.eps} \\
  \end{tabular} 
 \end{center}
 \caption{\label{f3}
  Cumulative distribution $N(<H)$ of the absolute magnitude $H$ for known Hobson family
  members. {\em Left panel}: All members included: the red and blue diamonds denote the
  two largest remnants (18777)~Hobson and (57738) 2001~UZ160, and filled circles show
  small members in the family. The gray line shows the power-law relation $N(<H)
  \propto 10^{\gamma H}$ for $\gamma=0.95$, which approximates the distribution of small
  members. {\em Right panel}: The second largest member, (57738) 2001~UZ160,
  is excluded from the distribution: the red diamond denotes (18777)~Hobson, and the filled
  circles are smaller fragments in the family. The power-law relation $N(<H)\propto 10^{\gamma H}$
  has now $\gamma=1.15$ (simply as an effect of removing the second largest body from the sample
  and linear-logarithmic scales on the axes). The upper abscissa in both panels provides an
  estimate of the size, assuming the geometric albedo $p_{\rm V}=0.2$ (a plausible value for
  S-type asteroids).}
\end{figure*}

Our search in this box zone resulted in $69$ objects that are clearly divided into two
populations: (i) $24$ dispersed asteroids filling the whole zone roughly uniformly, these are the
background population, and (ii) $45$ objects (including Hobson) that are tightly clustered about 
the origin, which represent the
Hobson family. Examples are shown in Figs.~\ref{f1} and \ref{f2}, where we show
projections onto a plane of secular angles $\Omega$ versus $\varpi$ and $\Omega$ versus
$I$, respectively. The compactness of the Hobson family in secular angles
(Fig.~\ref{f1}) is impressive: if we were to statistically characterize their
distribution, we would obtain $\Omega_{\rm fam} = 105.13^{+0.50}_{-0.96}$ in degrees and
$\varpi_{\rm fam} = 285.93^{+2.22}_{-1.83}$ in degrees. Compare these values to the dispersion of nearly an
order of magnitude larger in both angles of the $500$~kyr old Datura
family \citep[see Fig.~12 in][]{datura2017} and to the basically nonclustered values of
secular angles for $\leq 4.5$~Myr old Nele family \citep[see Fig.~5 in][]{cetal2018}.
There is hardly any doubt about the statistical significance of the Hobson family.
Just taking the data from Fig.~\ref{f1}, we may interpret the $24$ scattered background
objects as evidence of a background population density of $\sim 0.01$ per degree square.
In the $3^\circ\times 5^\circ$ vicinity of (18777)~Hobson, we have $45$ asteroids,
therefore the object density locally is $\sim 3.0$ per degree square. This argument does
not even take into account the significant clustering in all other osculating elements of our
initial search (Fig.~\ref{f2}). Additionally, we numerically integrated the nominal orbits
of all $69$ objects in the box zone of our search backward in time, reaching the $1$~Myr
epoch before the present. The orbits of all suggested Hobson members indicate a convergence of
the secular angles to (18777)~Hobson some $320$~kyr ago (Sec.~\ref{conv}), while the orbits
of the proposed $24$ background asteroids do not converge at any moment during our
test. The complete list of currently known Hobson members is provided in Table~\ref{tab_prop}
of Appendix~A.

\subsection{Size distribution} \label{sfd}
After we identified the currently known members of the Hobson family, we proceeded with
further analyses. We started with the absolute magnitude $H$ (or size $D$) distribution
of the Hobson members. As mentioned in Sec.~\ref{lfs}, accurate absolute magnitude values
have been obtained from dedicated observations of the two largest fragments (18777)~Hobson
and (57738) 2001~UZ160. Unfortunately, no data of comparable accuracy are available
for any of the other members. We therfore used absolute magnitudes from the {\tt MPCORB.DAT}
database. In addition to a random component in their uncertainty, \citet{petal2012} argued
that these values may have a systematic offset of $\simeq 0.25$~magnitude for their
respective $H$ range (Fig.~1 in that reference). If we had to map the magnitude
distribution to the size distribution, we may adopt the simplifying assumption of a
$p_{\rm V}=0.2$ geometric albedo based on S-type classification of the largest members
in the family (Sec.~\ref{lfs}).

Figure~\ref{f3} shows our results. We start with the left panel, which provides
the magnitude distribution of all members in the Hobson family. Starting from the third
largest fragment at $H\simeq 17.1$~magnitude, the spectrum of magnitudes among the
Hobson members steeply increases and may be approximated, at least in about $1$ magnitude interval
of $H$ values, with a power law $N(<H)\propto 10^{\gamma H}$ with
$\gamma\simeq 0.95$. Beyond magnitude $18,$ a strong incompleteness produces a bend
in the distribution, which obviously brings the question which features hold and which
are biased in our $H(<H)$ knowledge. This is in principle a difficult question, and
we answer it only approximately using information in \citet{hm2020}. These
authors developed an empirical approach to characterize the completion limit $H_{\rm lim}$
of the main belt population of asteroids as a function of semimajor axis. At $a
\simeq 2.56$~au, where the Hobson family is located, they obtained $H_{\rm lim}\simeq
17$ magnitude (with about $0.2$ magnitude uncertainty). This information suggests that
(i) the population of Hobson members may be completely known at the large end,
in particular, we may not be missing any fragment in the ``gap'' of the
$H(<H)$ distribution before the steep increase starts at about $17$~magnitude, and
(ii) the population beyond $17$~magnitude becomes incomplete, with the implication
that the true power-law exponent $\gamma$ is steeper than the one we find here
(Fig.~\ref{f3}).
With this information, we return to the magnitude distribution of the Hobson members
that should be known completely, namely the largest bodies. As we recalled in
Sec.~\ref{lfs}, this is represented by the pair of nearly equal-sized asteroids
(18777) Hobson and (57738) 2001~UZ160. In the next two sections, we verify
the mutual convergence of their orbits in the past and also confirm the convergence with other
smaller members in the family. This excludes the possibility that either of them
would be an interloper in the Hobson family. On the other hand, an
arrangement of fragments in magnitude distribution as shown in the left panel of
Fig.~\ref{f3} is not seen in any of the known families,%
\footnote{This issue puzzled already \citet{pv2009} who wondered if (18777) Hobson
 belongs to their proposed cluster around (57738) 2001~UZ160 (their Table~6).}
and it has yet to be
reproduced in computer simulations of family-formation events. We consider, for instance,
the most relevant set of such simulations in \citet{setal2017}. These authors
conducted a large suite of simulations, in which a $10$~km size parent asteroid
was hit by projectiles of various sizes between $\simeq 0.3$~km to $1.85$~km,
with various impact velocities between $3$ km~s$^{-1}$ to $7$ km~s$^{-1}$, and at
various impact angles. The situations covered basically all impact energy regimes, from
subcatastrophic cratering events to supercatastrophic disruptions
\citep[see additional results of][where effects of parent body rotation were
studied]{setal2019}. The mosaic
of all possible resulting size (or equivalently, magnitude) distributions of
fragments is shown in Fig.~1 of \citet{setal2017} or Fig.~2 of
\citet{setal2019}. None of them matches the observed distribution in the left
panel of Fig.~\ref{f3}. Either (i) some assumptions in the \citet{setal2017} and
\citet{setal2019} studies were not applicable to the Hobson family parent object
(possibly related to its internal structure or unsampled part of multidimensional
parametric space of the reported disruption models), or (ii) something else must
be taken into account.

If the first, we need to rerun simulations similar to \citet{setal2017} and
\citet{setal2019} with the emphasis on unsampled parametric combinations or so far
unused assumptions about the internal mechanical properties of the target body.
If the second, we need to seek a more unconventional solution. In this respect, we
note the results from \citet{petal2016}. These authors studied
small binary asteroids in the near-Earth, Hungaria, and inner main-belt
populations. For the inner main-belt group, \citet{petal2016} confirmed that
about $15\pm 4$\% of $D<15$~km asteroids are binary, at least half
of which are systems with a close-by spin-orbit synchronized satellite on a nearly
circular orbit. The typical size ratio of the satellite and primary in these
binary systems is between $0.2$ and $0.4$, the typical satellite distance from
the primary is between $2$ and $3$ (referred to as the primary size), and the typical
orbital periods of the satellite are between $15$~hr to $30$~hr. Given their
significant abundance in the population, it is conceivable that the parent objects
of some small asteroid families in the inner main-belt would be binaries.
Depending on the geometry of the impact, either primary or the satellite
(or possibly both) may be involved in the collision and fragmentation. It is
therefore possible, as an example, that the primary component in the
binary collisionally disrupts and feeds the family with its fragments,
leaving the former satellite largely intact.

Translated into the reality of the Hobson family, we might assume that (57738) 
2001~UZ160 is the former satellite of the parent binary, for example. If the 
geometry of the family-forming impact allows it, this member may still
record the properties of the satellite, including its size of $\simeq 2.5$~km
and the synchronized rotation period of $20.5$~hr. These values are
plausible, and they would imply an expected size of the broken primary
in the $6$~km to $12$~km range. The characteristic initial satellite
orbit would have $15$~km to $30$~km radius. Because the entire fragmentation process
concerns the former primary in the binary system, the right panel of
Fig.~\ref{f3} would be relevant for comparison with computer simulations
of small-asteroid disruptions such as in \citet{setal2017} or \citet{setal2019}.
Asteroid (18777)~Hobson would be the largest remnant, and the suite of
kilometer-sized and smaller asteroids would be the small-size tail in
the distribution. We note that the power-law approximation $N(<H)\propto
10^{\gamma H}$ of the tail population would have $\gamma=1.15$ now. Observational
limitations imply that the true (unbiased) distribution would have $\gamma > 1.2$
(say). The $2$~magnitude gap between (18777)~Hobson and the foot of the tail translates
into a factor $10^{-0.4}\simeq 2.5$ in size (independent of the surface albedo value).
These are then the parameters to be matched by the results from computer
simulations of asteroid disruptions (together with the proof that the
secondary component in the parent binary remains intact).

In Sec.~\ref{sfd_mod} we explore both possibilities using a smoothed-particle
hydrodynamics (SPH) code for the fragmentation phase, followed with an N-body integrator
for subsequent gravitational reaccumulation (thence the SPH/N-body numerical
approach) based on the methods presented in \citet{setal2017} and \citet{setal2019}
(see also Appendix~B). Since the works of \citet{mi2001,mi2003}, this approach became
the default tool for modeling the asteroid family formation. In particular,
results form the SPH/N-body simulations allow us to compare the resulting size distribution
of the synthetic Hobson family with the data shown above. We also obtain a prediction
of the characteristic dispersal speeds of the fragments with respect to the
largest remaining fragment, and these may be compared with results from our
convergence experiments in Secs.~\ref{conv} and \ref{pair}.
\begin{figure*}[t]
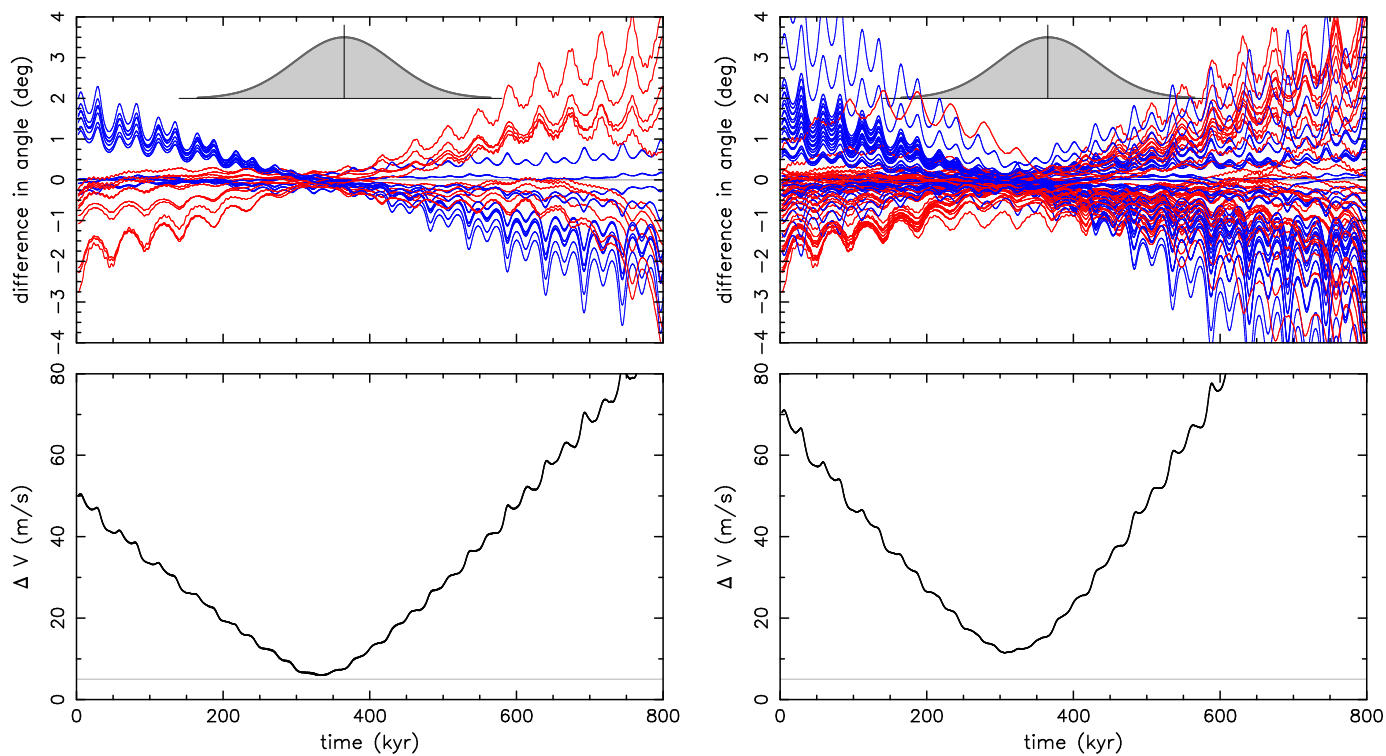

 \begin{center}
  \begin{tabular}{cc}   
   \includegraphics[width=0.48\textwidth]{f4a.eps} &
   \includegraphics[width=0.48\textwidth]{f4b.eps} \\
  \end{tabular} 
 \end{center}
 \caption{\label{f4}
  {\em Top}: Convergence of the mean values of secular angles, longitude of node in red and
  longitude of perihelion in blue, for Hobson family members in the base model with
planetary perturbations alone. The time at abscissa
  extends to the past, and the ordinate shows the difference of the respective angle with respect
  to the orbit of (18777)~Hobson. The reference frame is defined by the invariable plane of
  the Solar System. {\em Bottom}: Value of target function $\Delta V$
  from Eq.~(\ref{tf}) computed for the orbits shown in the top panel. The gray horizontal
  line indicates $5$ m~s$^{-1}$, the estimated escape velocity from the family parent
  asteroid (Sec.~\ref{sfd_mod}). {\em Left panels} show 12 members with $H\leq 17.75$, and the {\em
  right panels} show all Hobson members on multi-opposition orbits. The gray Gaussian curve in 
  the top panels schematically indicates the solution $365\pm
  67$~kyr for the age of the Hobson family from \citet{rp2017}. }
\end{figure*}

\subsection{Nominal convergence of secular angles} \label{conv}
We now return to the issue of a past orbital convergence of members in the Hobson family.
As discussed in Sec.~\ref{intr}, in our context, we aim at shrinking the differences in
the secular angles $\Omega$ and $\varpi$ to near-zero values for as many orbits in the
family as possible. The present-day $\simeq 1^\circ$ and $\simeq 2^\circ$ dispersion
values in $\Omega$ and $\varpi$ may already look very small, but in fact, they are
not. Using the Gauss equations of the perturbation calculus \citep[e.g.,][]{nv2006},
it is easily clear that they still correspond to nearly $60$ m~s$^{-1}$ dispersal
in velocities (using Eq.~\ref{tf} below). These values are expected in the largest
families with $>100$~km size parent objects, but far lower values are seen in
tighter families \citep[as an example, see][]{karin2006}. This observation
motivates a need of further convergence of secular angles in the past.
\citet{rp2017,rp2018} have already shown that this is possible for up to six members in
the Hobson family. Here we take up their effort and extend it to all currently
known members in the family (excluding only the poorly determined case of the six
single-opposition members 2014~JH120, 2014~OJ66, 2020~JM31, 2020~KP36, 2020~OY50,
and 2021~MO5; Table~\ref{tab_prop}).

While the behavior of the secular angles of all propagated orbits is the basis of
the convergence test, \cite{nv2006} found it interesting to combine this multidimensional
information into a single target parameter,
\begin{equation}
 \Delta V = na \sqrt{\left(\sin I \Delta \Omega\right)^2 + 0.5
   \left(e\Delta \varpi\right)^2}\; , \label{tf}
\end{equation}
where $na\simeq 18.6$~km~s$^{-1}$ is the characteristic orbital velocity of the Hobson members,
$e$ and $\sin I$ are the orbital eccentricity and inclination (we may use the values of
(18777)~Hobson), and $\Delta \Omega$ and $\Delta \varpi$ are dispersal
values of longitude of node and perihelion. These quantities are defined as $\left(\Delta
\Omega\right)^2 = \sum_{ij} \left(\Delta \Omega_{ij} \right)^2/N$, where $\Delta
\Omega_{ij}$ are simple differences in nodal longitudes of $i$th and $j$th objects,
and $N$ is the number of pair combinations between asteroids tested (and similarly
for perihelia). In this way, the target function $\Delta V$ has a dimension of velocity
and in a statistical sense, approximates the magnitude of the dispersal velocity among the orbits
in the cluster. In spite of using only $\Omega$ and $\varpi$ angles, $\Delta V$
depends on all three components of the velocity vectors and represents an average
over the position in heliocentric orbit (true anomaly) in which different orbits are
compared. Observing structure of the Gauss equations, \citet{rp2018} proposed an
alternative target function, 
\begin{equation}
 \Delta V_{\rm Z} = na \sqrt{\left(\Delta I\right)^2 +
   \left(\sin I\Delta \Omega\right)^2}\; , \label{tf1}
\end{equation}
where $\left(\Delta I\right)^2$ is now the dispersion of the inclination values among the
propagated orbits. While $\Delta V_{\rm Z}$ again represents a statistical mean over
the phase of the heliocentric motion, it provides information about the normal component
$V_{\rm Z}$ to the orbital plane, statistically averaged over the pair identification of
orbits in the cluster. Therefore $\Delta V_{\rm Z}$ conveniently isolates the information
about this normal velocity component, but does not tell us anything about the two
in-orbit components. The diagnostic advantage of either $\Delta V$ or $\Delta V_{\rm Z}$
consists of their dependence on at least one of the secular angles. This is because
these angles drift secularly, in contrast to the semimajor axis, eccentricity, or inclination
values, which only oscillate with terms of various periods (both shorter and longer than
the possible age of the family).
\begin{figure*}[t]
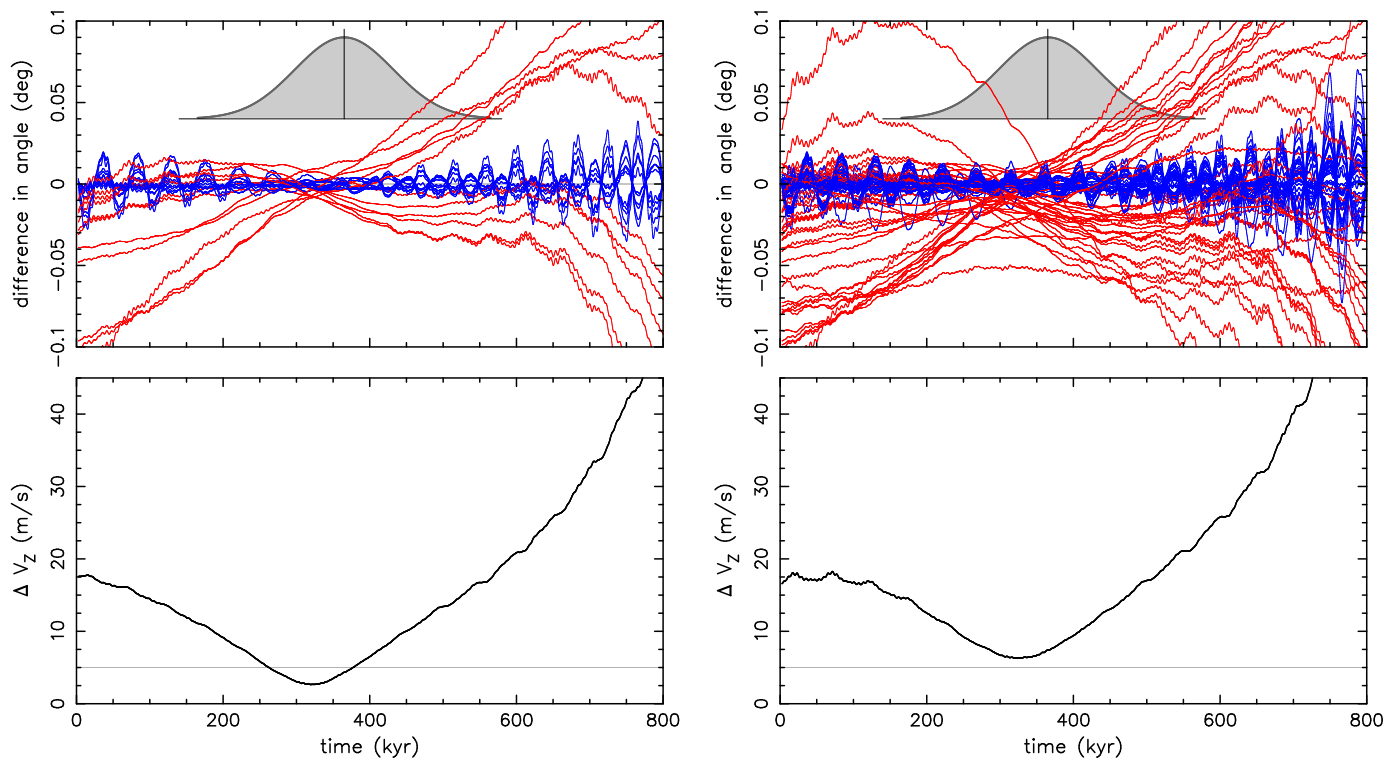

 \begin{center}
  \begin{tabular}{cc}   
   \includegraphics[width=0.48\textwidth]{f5a.eps} &
   \includegraphics[width=0.48\textwidth]{f5b.eps} \\
  \end{tabular} 
 \end{center}
 \caption{\label{f5}
  {\em Top}: Convergence of the mean values of inclination (blue curves) and longitude of node
  multiplied with a sine of inclination (red curves) for Hobson family members in
  the base model with planetary perturbations alone. The time
  at abscissa extends to the past, and the ordinate shows difference of the respective variable
  with respect to the orbit of (18777)~Hobson. The reference frame is defined by the invariable plane
  of the Solar System. {\em Bottom}: The value of the target function $\Delta V_{\rm Z}$
  from Eq.~(\ref{tf1}) computed from the orbits shown in the top panel. The gray horizontal
  line indicates $5$ m~s$^{-1}$, the estimated escape velocity from the family parent
  asteroid (Sec.~\ref{sfd_mod}). {\em Left panels} show 12 members with $H\leq 17.75$, and the {\em
  right panels} show all Hobson members on multi-opposition orbits.
  The gray Gaussian curve schematically indicates the solution $365\pm 67$~kyr for the
  age of the Hobson family from \citet{rp2017}. }
\end{figure*}

Ideally, $\Delta V$ or $\Delta V_{\rm Z}$ thus exhibit a clear minimum at the
epoch of the family origin. In an ideal world, the result would be easy: one simulation would
be enough to provide the exact epoch of the target function minimum (i.e., family origin),
and its value would be the velocity dispersion. The real world is more complex because
there are many more degrees of freedom to be considered. They have to do with our ability
to reconstruct the orbital architecture of the family members in the past (typically
hundreds of thousands of years for detected very young families). This is influenced by
two aspects: (i) the initial orbital conditions of family-member asteroids are not
known accurately even at the current epoch, and (ii) the mathematical model, which allows us
to reconstruct the past states of the family orbits, may not be completely constrained
or exhibit chaoticity. Both effects would require that every asteroid in the family
is represented by a variety of clones for both (i) and (ii), and a huge number of
possible clone mutual identifications would have to be considered. This is because
each of these possible clone configurations would have its own time dependence on the
target functions $\Delta V$ or $\Delta V_{\rm Z}$ in the past. Despite their global
similarity, that is, most of them would reach a minimum at some epoch, they would not be
identical. Shifts in (i) epochs of minima, and in (ii) minimum values would be inevitable.
A statistical criterion would be needed to describe the results.

Starting with the works of \citet{datura2006} and \citet{nv2006}, researchers mostly
used one possible variant. With a {\em \textup{predefined}} tolerance value of $\Delta V$ magnitude,
for instance, the estimated escape velocity from the parent objects of the family, they
considered all possible $\Delta V$ functions, individual to clone configurations, that
met the criterion and characterized them statistically. The goal of this approach was
primarily the determination of the family age and its uncertainty, while the velocity
dispersal at origin was only {\em \textup{assumed}}. We did not follow this way. We are less
interested in the age of the Hobson family, but we wish to learn something about the
initial dispersal velocity of the observed fragments. We approached this goal using the
simplest method, namely by using (i) just the nominal (best-fit) orbits of the family
members, and (ii) the simplest force model for the orbital propagation. In (ii) we
consider only the gravitational effects from the Sun, planets, and largest asteroids
(Ceres, Pallas, and Vesta). We do not include the effects of thermal accelerations
known as the Yarkovsky effect. This brings us back to the single, ``ideal world''
simulation mentioned above. However, the difference is in the interpretation of the
results. We wish to determine the minimum of $\Delta V$ or $\Delta V_{\rm Z}$ that can be reached
in the simulation, and we consider it to be {\em \textup{the upper limit}} of the initial
velocity dispersal of the family fragments. It is obvious that adding more degrees
of freedom, represented by the clone variants of the objects (especially those related
to the thermal accelerations), will allow us to decrease the minima of $\Delta V$ or
$\Delta V_{\rm Z}$. Quite likely, the true history of the family will be among these
improved solutions. This is because it would appear strange that the hugely
simplified base model would be better than truth.

We used a well tested {\tt swift\_mvs} orbit propagation package%
\footnote{\url{http://www.boulder.swri.edu/~hal/swift.html}}
for our runs. Planetary
initial conditions were taken from the JPL ephemerides, and asteroid orbits (including Ceres,
Pallas, and Vesta) were taken from the {\tt AstDyS} site.%
\footnote{\url{https://newton.spacedys.com/astdys/}}
All data were referred to the initial epoch MJD 59200.0, and orbital velocities were reversed
to perform backward-in-time integration. We used a $\text{three-day}$~ time step and output the state
vectors of all bodies every $\text{five}$~years. The longest epoch reached in our simulation is
$1$~Myr. We performed two simulations: (i) the base model that only included planetary
perturbations, and (ii) the extended model that also contained perturbations by
Ceres, Pallas, and Vesta. This allowed us to estimate the role of the massive
objects in the main belt on the Hobson family history.
Before we report the results, we computed mean orbital elements of the
Hobson members by eliminating short-period terms with peroids shorter than $500$~yr.
This is because our output is too sparse, and additionally, numerical integration
is subject to effects of dynamical chaos, to resolve family formation conditions in
the phase of motion about the Sun.%
\footnote{We also performed tests in which we used osculating elements and a very high
 temporal cadence of every time step of the integrator in evaluating the target
 functions (\ref{tf}) and (\ref{tf1}). The resulting values of $\Delta V$, for instance,
 exhibit high-frequency oscillations about the signal shown in the lower panels of Fig.~\ref{f4},
 reaching slightly lower minimum values. Generally, our reported $\Delta V(t)$,
 and similarly $\Delta V_{\rm Z}(t)$, approximate the lower bound of these value rather well, however.
 A detailed study of the high-frequency signal patterns from our tests, with the goal of
 refining the Hobson family age, is not a goal of this study, however, and it is postponed
 to a future work.} 

The upper panels in Fig.~\ref{f4} show the behavior of the secular angles, the longitude of
node (in red), and longitude of perihelion (in blue), referred to the value of
(18777)~Hobson in the base simulation. The left panels show the $12$ largest members
in the family with $H\leq 17.75$~magnitude, and the right panels show all $39$ multi-opposition
members (see Table~\ref{tab_prop}). Presumably, the orbits of the additional set
of small asteroids in the right panels are less constrained and potentially subject to
stronger perturbations from the Yarkovsky effect. The bottom panels then summarize
the data from the top panel into the numerical value of the $\Delta V$ target function
from Eq.~(\ref{tf}). We first discuss the results for $12$ large asteroids at the
left. We note an excellent degree of simultaneous convergence of both nodes and perihelia
some $330$~kyr ago. This agrees well with similar results of \citet{rp2017}, but
now for more than twice as many objects. In response, the $\Delta V$ target function
has a sharp minimum at the same epoch, reaching $5.9$ m~s$^{-1}$ at best. This value
is impressively close to the estimated escape velocity from the parent body of the
Hobson family (Sec.~\ref{sfd_mod}). When less accurate orbits of small fragments are
included in the simulation, now a total of $39$ orbits in right panels, the
convergence becomes slightly defocused, and therefore the minimum $\Delta V$ value
is slightly higher than $11.5$ m~s$^{-1}$. This was expected, but the optimum
$\Delta V$ values are still quite low. This witnesses the rather small dispersal
velocity field down to fragments of $\simeq 0.5$~km. The minimum now
corresponds to $307$~kyr epoch, but the difference is well within the formal uncertainty
value stated in \citet{rp2017}.

The results do not change much when perturbations from Ceres, Pallas, and Vesta are included
in the extended simulations. For instance, the minimum values of the $\Delta V$ target
function are (i) $9.3$ m~s$^{-1}$ when the $12$ largest fragments are used (as in
the left panels of Fig.~\ref{f4}), and (ii) $16.2$ m~s$^{-1}$ when all $39$ fragments
are used. This is only slightly worse than in the base simulation. Unless very close
encounters, the effects of the three most massive objects in the main belt (Ceres
in particular) are moderate on the few hundred kyr timescale of interest.

Panels on Fig.~\ref{f5} show the results we obtained using the $\Delta V_{\rm Z}$ target function
(bottom) and the relevant angular differences with respect to the orbit of
(18777)~Hobson (see Eq.~\ref{tf1}): (i) inclination (in blue), and (ii) longitude
of node multiplied with sine of inclination (in red). As expected, only the latter
component shows a converging pattern, while the inclination oscillates. However,
a sign of convergence for the inclination is expressed by a decrease in the oscillation
amplitude. The convergence feature may look less impressive than in Fig.~\ref{f4},
but note the ordinate scale of both figures. The bottom panels show the behavior of
the $\Delta V_{\rm Z}$ target function. Its minimum occurs consistently at about
$320$~kyr ago, and the minimum values are (i) $2.6$ m~s$^{-1}$ when the $12$ largest
fragments are taken into account (left), and (ii) $6.4$ m~s$^{-1}$ when all $38$
fragments are taken into account (right). Interestingly, these values are not
far from a factor $\sqrt{3}$ smaller than the corresponding minimum values of
$\Delta V$ target function. This may indicate near isotropy of the ejection field
of the Hobson fragments at origin.

Finally, we recall that our definition of the target function $\Delta V$ in
Eq.~(\ref{tf}) contains root-mean-square $\Delta\Omega$ and $\Delta\varpi$ values
that were computed over all pair combinations of the orbits \citep[definition introduced
by][and similarly for $\Delta V_{\rm Z}$ in Eq.~(\ref{tf1})]{nv2006}. In this way, (i)
there is no preferred orbit, but (ii) the $\Delta V$ and $\Delta V_{\rm Z}$ values
may be conservatively high by an outlying contribution from distant orbits of small
fragments in the family. Alternatively, a priori,  a reference orbit in
the family may be selected and a modified target function $\Delta V'$ and 
$\Delta V'_{\rm Z}$ introduced, in which the $\Delta\Omega$ and $\Delta\varpi$ are 
the root-mean-square values of the difference with respect to the reference orbit 
alone. If the reference orbit resides
well within the center of the family, these values may be lower. For the sake of
a test, we constructed these alternative options $\Delta V'$ and $\Delta V'_{\rm Z}$
of the target function using two different reference orbits: (i) (18777) Hobson,
and (ii) (57738) 2001~UZ160.

When the orbit of (18777) Hobson is chosen as a reference, $\Delta V'$ reaches a minimum of
about $4.4$ m~s$^{-1}$ when orbits of $H<17.75$ members are used, and $8.7$ m~s$^{-1}$
when all members on multi-opposition orbits are used. These values are slightly
lower than the minima in the bottom panels of Fig.~\ref{f4}, $5.9$ m~s$^{-1}$ and
$11.5$ m~s$^{-1}$ , respectively. This confirms that the orbit of (18777) Hobson 
suitably lies in the center of the family. Interestingly, when the orbit of
(57738) 2001~UZ160 is chosen as a reference, $\Delta V'$ reaches a minimum of
about $8.9$ m~s$^{-1}$ when orbits of $H<17.75$ members are used, and $12.2$ m~s$^{-1}$
when all members on multi-opposition orbits are used. These values are higher,
implying that the nominal orbit of (57738) 2001~UZ160 is slightly offset from the
true center of the family. This may be an interesting indication in support of our
model in which the parent object of the Hobson family is a binary (see discussion
in Sec.~\ref{sfd_mod}). However, the differences are small and should not be
overstated. A more thorough convergence analysis, in which geometrical and
Yarkovsky clones of the family members are taken into account, could resolve
this issue in the future.

\subsection{(18777) Hobson and (57738) 2001~UZ160 as a pair} \label{pair}
In Sec.~\ref{sfd} we discussed the special status of the two largest members
in the Hobson family, asteroids (18777) Hobson and (57738) 2001~UZ160. In our
formation model from a parent binary, the former is the largest fragment from the family-forming
event, while the
latter is the surviving satellite of the proto-binary. In the previous section,
we verified that the heliocentric secular angles of both asteroids tend to
mutually converge at approximately the same epoch as all other orbits. Here we
strengthen their relation by proving the possibility of a past convergence of their
full Cartesian state vectors. This technique is only possible for two heliocentric
orbits and was first used by \citet{vn2008} for an analysis of asteroid pairs.
\begin{figure}[t]
 \begin{center}
  \includegraphics[width=0.49\textwidth]{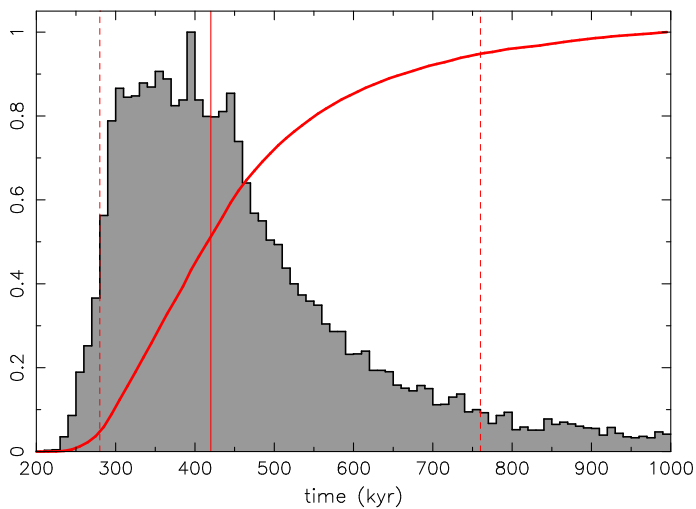}
 \end{center}
 \caption{\label{f6}
  Distribution of convergent configurations for clones of (18777)~Hobson and
  (57738) 2001~UZ160 in Cartesian space with the following criteria: (i) physical
  distance $\leq 10000$~km, and (ii) relative velocity $\leq 4$ m~s$^{-1}$. The
  gray histogram is a differential distribution using $10$~kyr bins normalized to
  maximum convergent cases in a bin. The red line is a cumulative distribution.
  Time at abscissa extends to the past. Red vertical lines delimit median convergent time
  (solid), $5$\% and $95$\% levels of the cumulative distribution (dashed).}
\end{figure}

We thus use numerical propagation of heliocentric orbits of (18777)~Hobson and
(57738) 2001~UZ160 backward in time to prove their close approach in physical space
at a very low relative velocity. These conditions once occurred when the Hobson family
was formed. In particular, the zone of family-formation in our model has a characteristic
scale of the proto-binary, for instance, some $15$ to $30$~km. The expected relative velocity has
a smaller contribution from the orbital motion of the binary components (some $2$ m~s$^{-1}$
or less) and possibly larger component from the momentum imparted to the primary
by the impactor (some $4$ m~s$^{-1}$ or so). Ideally, these values set the quantitative
measure of state-vectors proximity we would like to achieve. While actually possible in optimum
cases \citep[see, e.g.,][who analyzed the $\simeq 16$~kyr old pair of asteroids (6070) Rheinland and
(54827) 2001~NQ8 on very stable orbits]{v6070}, these limits appear too ambitious for our
target. This is because of their quite old age (between $300$ and $400$~kyr, Sec.~\ref{conv})
and much less dynamically stable orbits \citep[see, e.g.,][who reported the chaotic nature of the Hobson
orbit, and our Appendix~A]{rp2016, rp2017}. Additionally, the two sources of uncertainty in
the reconstruction of the past asteroid state discussed in Sec.~\ref{conv} still hold. These
are due to the uncertainty in the initial conditions at the present epoch and to the incompleteness of
the orbit-propagation model due to unknown parameters of the thermal accelerations. In the
previous section, we did not take these effects into account. In contrast, we
need to include both effects here by considering a multitude of statistically equivalent clone
variants for each of the two asteroids. All these limiting aspects caused us to soften the
convergence requirements in our simulations: we assumed (i) a $10000$~km
physical distance of the clones (about $\text{five}$ times the Hill radius of the proto-binary of
the Hobson family), and (ii) a  $4$ m~s$^{-1}$ relative velocity. When any two clones met
these conditions, we considered them as a successful convergent situation. The best solutions 
we achieved bring the clones to a distance of several hundred kilometers
and to a relative velocity lower than $1$ m~s$^{-1}$, but given the limited number of the
clones we can afford, there are only few such solutions. To have enough statistical
information, we continued to assume the weaker convergence conditions.

The orbital uncertainty at the initial epoch of our simulation, MJD 59200.0, was
taken into account by representing each of the two asteroids by $5000$ clones. These
were generated using multidimensional Gaussian statistics in the space of
the equinoctic orbital elements provided by the {\tt AstDyS} website. The method takes
into account all correlations of orbital elements, which are not severe
(the largest occurs between the semimajor axis and the longitude in orbit), however. Additionally,
each of the clones was assigned a random value of thermal accelerations \citep[the
Yarkovsky effect, e.g.,][]{vetal2015}. Because only the rotation period
is known for (18777) Hobson and (57738) 2001~UZ160, we restricted ourselves to
the simplest approach to model the thermal effects \citep[see, e.g.,][]{nv2006,vn2008}.
We applied simulated transverse acceleration in the orbital dynamics of clones, which
resulted in the secular semimajor axis drift $da/dt$ predicted by the Yarkovsky effect.
These values have a uniform distribution in the range $(-(da/dt)_{\rm max},(da/dt)_{\rm max})$,
where $(da/dt)_{\rm max} = 3\times 10^{-4}/D$ au~Myr$^{-1}$ \citep[e.g.,][]{vetal2015},
with $D$ the estimated asteroid size in kilometers. We took the values derived in
Sec.~\ref{lfs}. Our dynamical model contains perturbations from all planets, and we
also took into account effects of the most massive objects in the asteroid belt: Ceres,
Pallas, and Vesta. Perturbations of Ceres, in particular, have been found to be non-negligible in
the Hobson zone by \citet{rp2016,rp2017}. We used a short integration time-step of $2$~days
and every $3$~years, which is slightly shorter than the orbital period of the asteroids in the Hobson family,
and we evaluated the distance and relative velocity of all possible $5000^2=25\times 10^6$
combinations of clones. A shorter frequency of the testing would clearly be
better, but is more demanding in CPU time. The simulation was pursued until a million-year
epoch in the past. As above, we used the {\tt swift\_mvs} software to carry out the
simulations efficiently.

The results from our simulation are shown in Fig.~\ref{f6}, where we plot the
statistical distribution of convergent clone combinations in the past. The histogram
showing the differential distribution has $10$~kyr bins
\citep[compare with Fig.~7a in][where similar results were obtained from fewer clones and
slightly different convergence criteria]{petal2018}. Taken at a face value,
the median, and $5$ to $95$\% confidence limits of the cumulative distribution
of the Hobson-2001~UZ160 age would be $420^{+340}_{-140}$~kyr. This is both (i)
shifted in median and (ii) wider in spread than the nominal solution $365\pm 67$~kyr
from \citet{rp2017} \citep[the median displacement is less of a problem because it is still
within the formal uncertainty of the][solution]{rp2017}. However, this is expected
because our simulation is
much more intensive.  \citet{rp2017} only considered nominal orbits of
a few Hobson members and disregarded the effect of the thermal accelerations in their
dynamics. Especially the latter produce the long-age tail in our solution and allow
a far wider range in the possible age of the Hobson-2001~UZ160 pair. As a result,
the solution of \citet{rp2017} is a subset of our age solution from the pair of the
two largest members in the Hobson family. On the other hand, the strength of the
\citet{rp2017} age consists of taking more than two orbits into account (this
is even far stronger in our Sec.~\ref{conv}). Unfortunately, it is not
possible to apply our Cartesian-space convergence method to more than two orbits.
However, it is obvious that considering more pairs of members in the Hobson family
would delimit its age more strictly and to a narrower interval of values. This goal is
beyond the scope of this paper, however. We are content here with a solid
justification of both (18777) Hobson and (57738) 2001~UZ160 as true members in the
family, and as explained above, setting the enveloping range for the Hobson
family age suffices.
\begin{figure}
\centering
\includegraphics[width=0.49\textwidth]{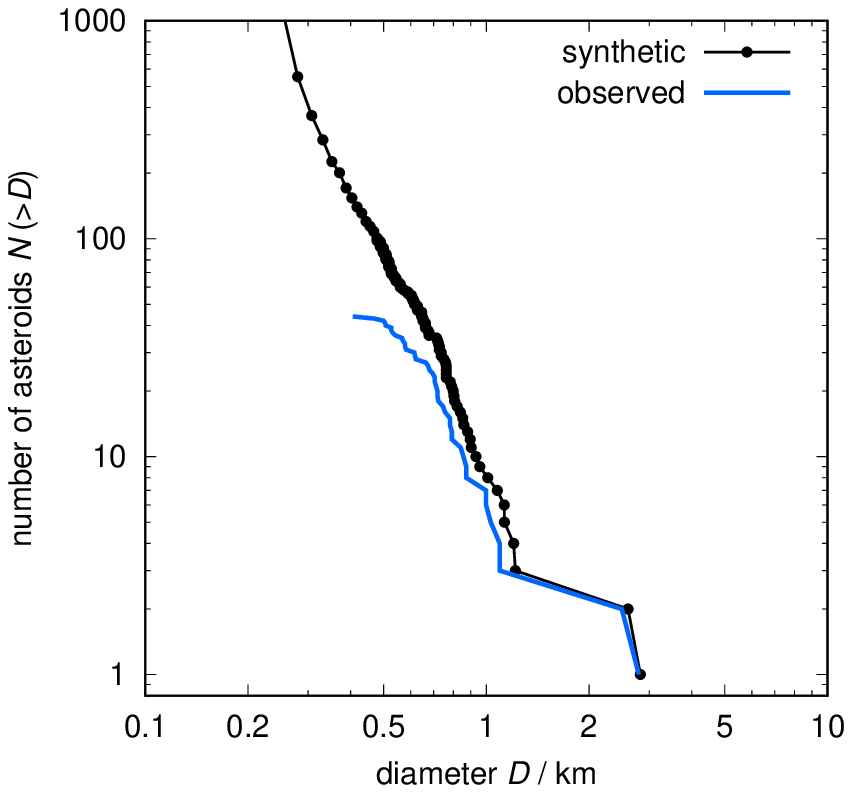}
\includegraphics[width=0.49\textwidth]{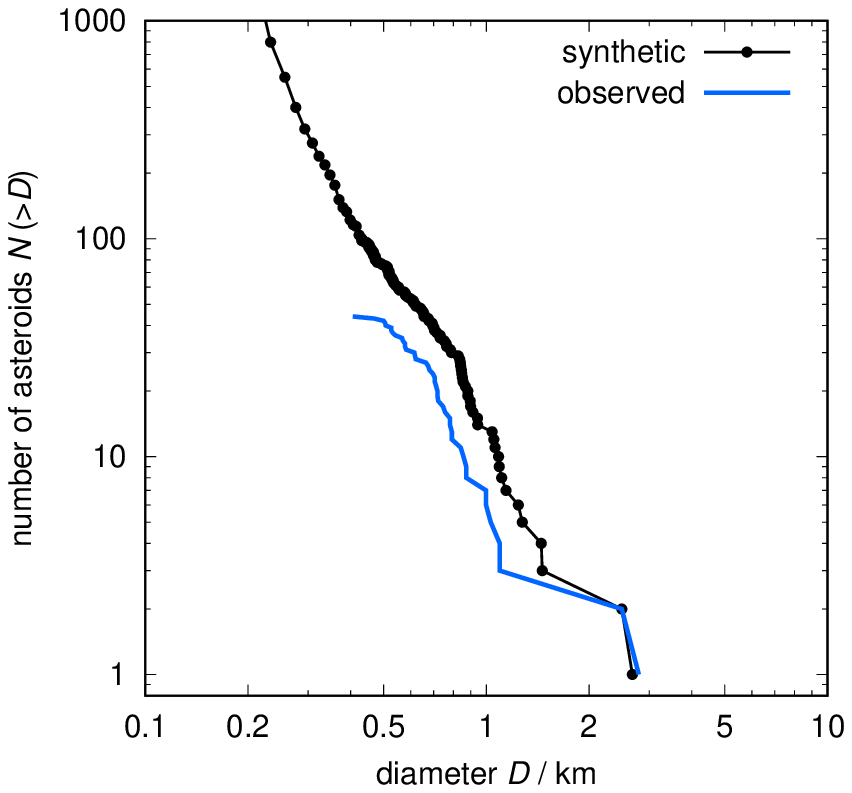}
\caption{\label{f7}
 Cumulative size-frequency distributions $N({>}D)$ of the fragments produced in
 the simulation where the parent object was a single asteroid ({\em top}) and
 a binary system ({\em bottom}). The observed Hobson-family data are shown
 with a blue line (assuming geometric albedo $p_{\rm V}=0.2$ as in Fig.~\ref{f3}).
 A slight shift in the populations of kilometer-size and smaller fragments may
 be partly explained by the incompleteness of the observed family.}
\end{figure}

\section{Numerical model of the Hobson family formation} \label{sfd_mod}
In order to understand formation of the Hobson family, we performed SPH/N-body
simulations of single as well as binary asteroid breakups. Binarity is an important
novel aspect compared to the previous models because the initial shock wave produced by
the impactor in the target cannot propagate to the secondary, but its mass contributes
to the total gravity \citep[e.g.,][]{Rozehnal_2016MNRAS.462.2319R}. If it survives intact,
the secondary appears as one of large remnants in the newly formed family. Therefore the
presence of a nearly preserved secondary
may potentially help to explain the peculiar size-frequency distribution (SFD) of the
Hobson family, which contains two similarly sized bodies (see also Sec.~\ref{sfd}).
\begin{figure*}
\centering
\newdimen\tmpdim\tmpdim=4.3cm
\begin{tabular}{cc@{\kern.05cm}c@{\kern.05cm}c@{\kern.05cm}c}
\raise1.0cm\hbox{\rotatebox{90}{fragmentation}} &
\includegraphics[width=\tmpdim]{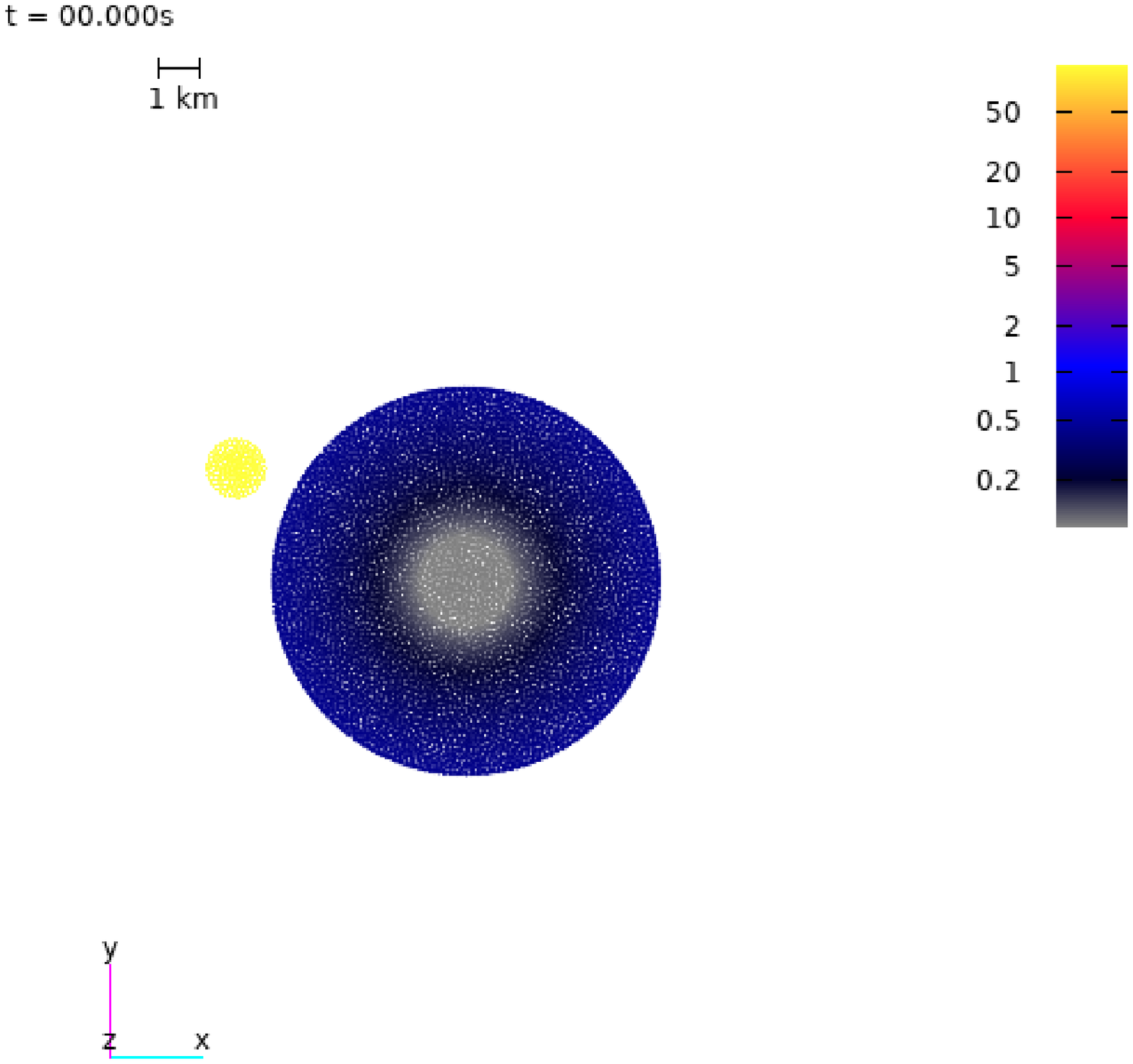} &
\includegraphics[width=\tmpdim]{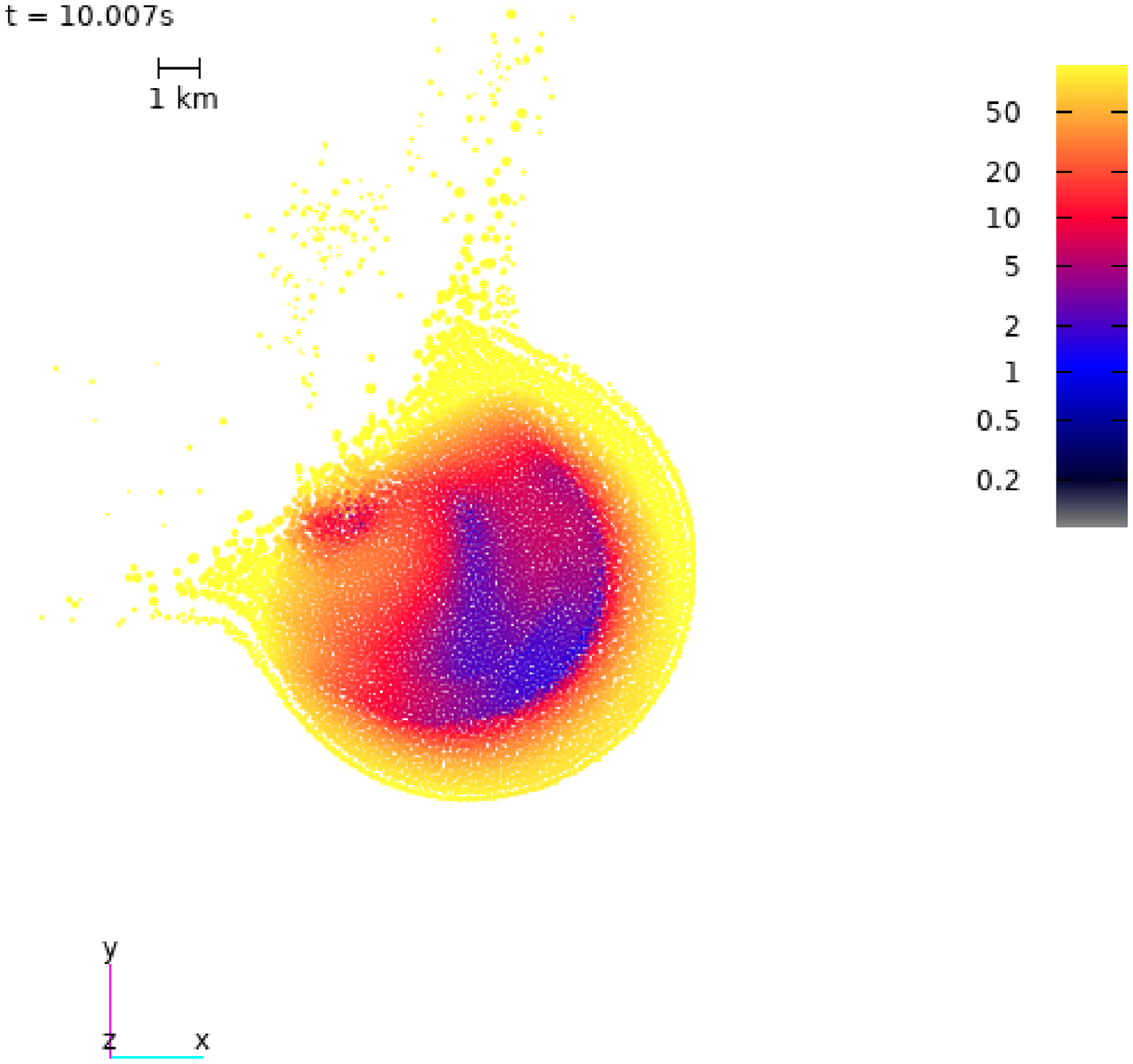} &
\includegraphics[width=\tmpdim]{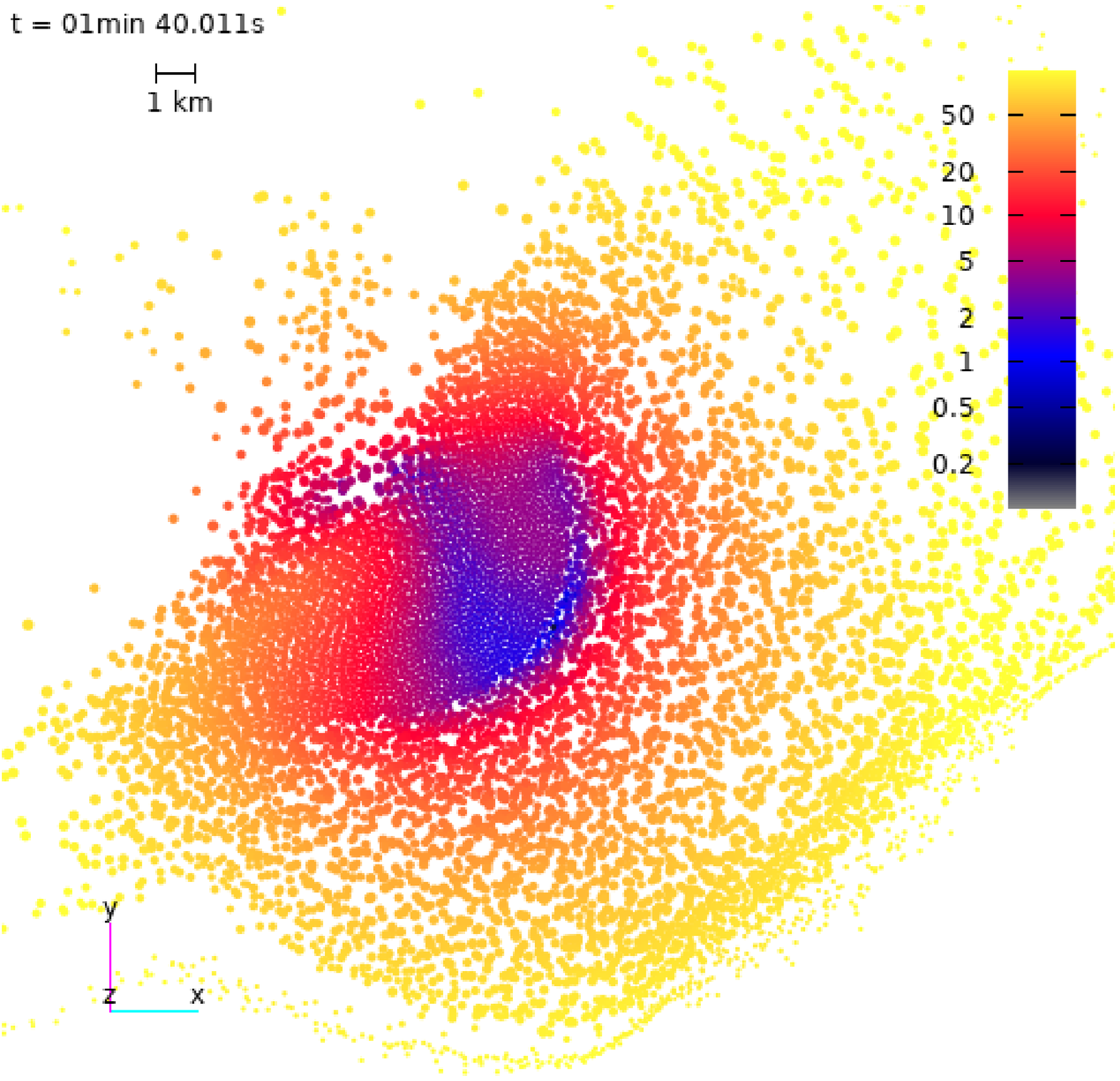} &
\includegraphics[width=\tmpdim]{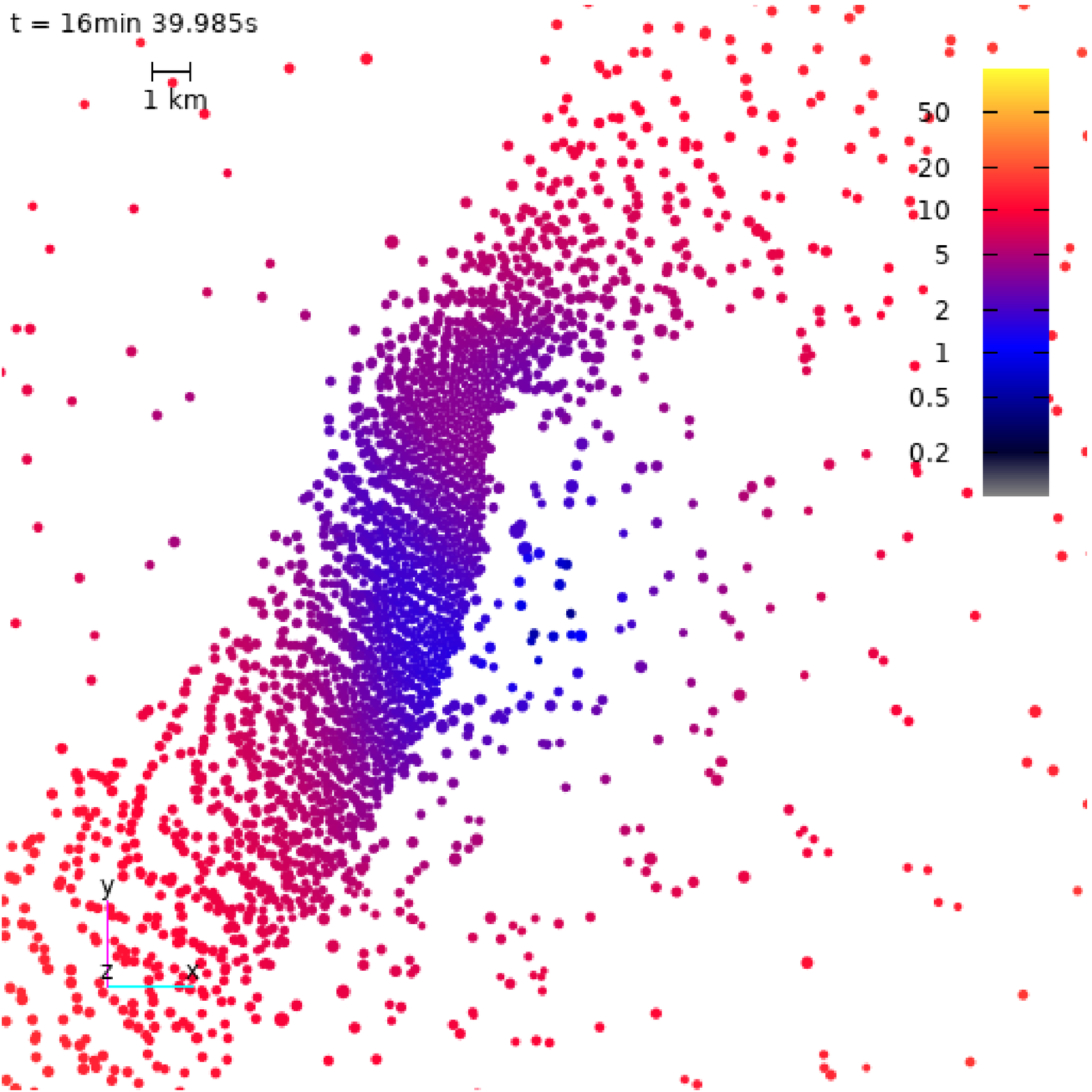} \\[-0.07cm]
\raise0.9cm\hbox{\rotatebox{90}{reaccumulation}} &
\includegraphics[width=\tmpdim]{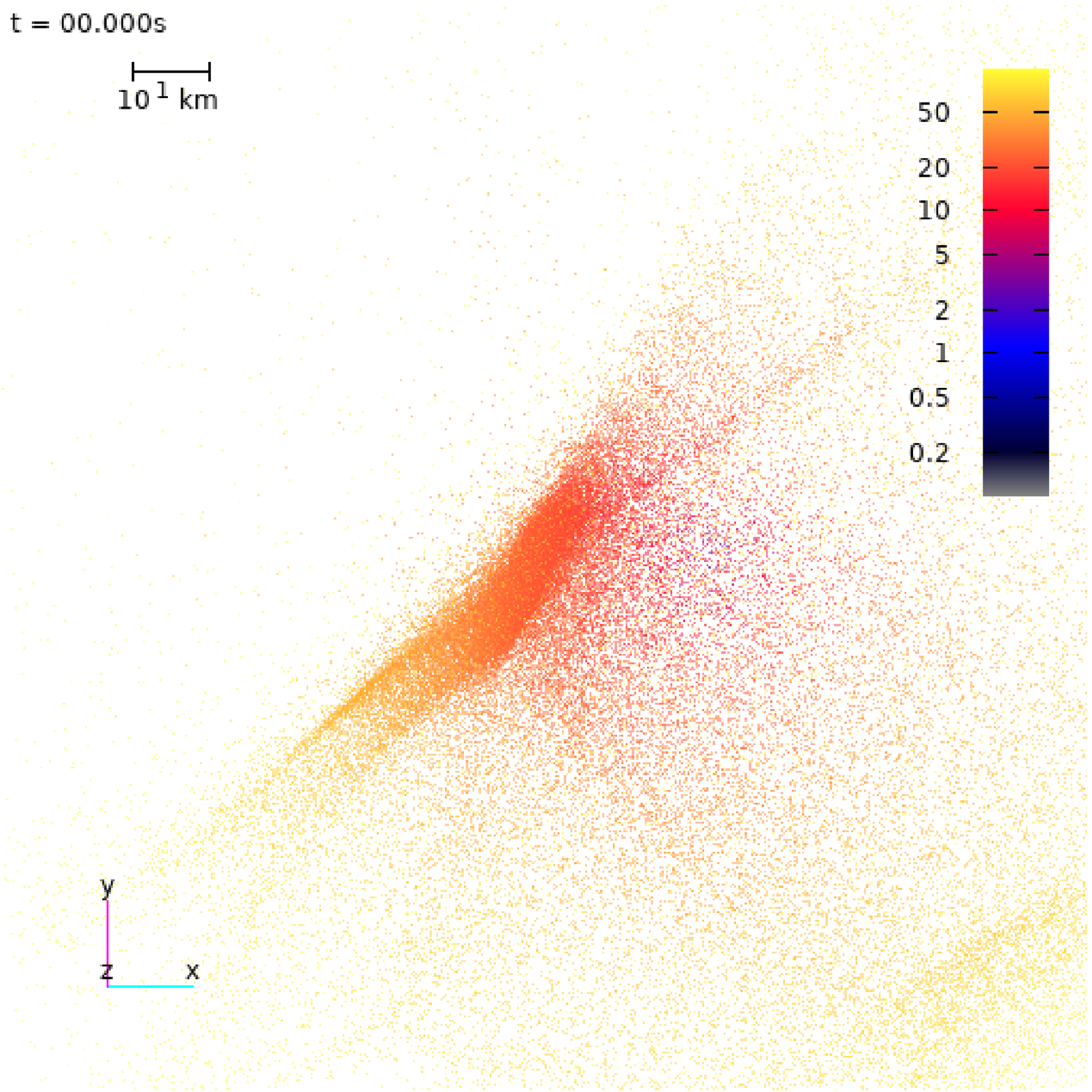} &
\includegraphics[width=\tmpdim]{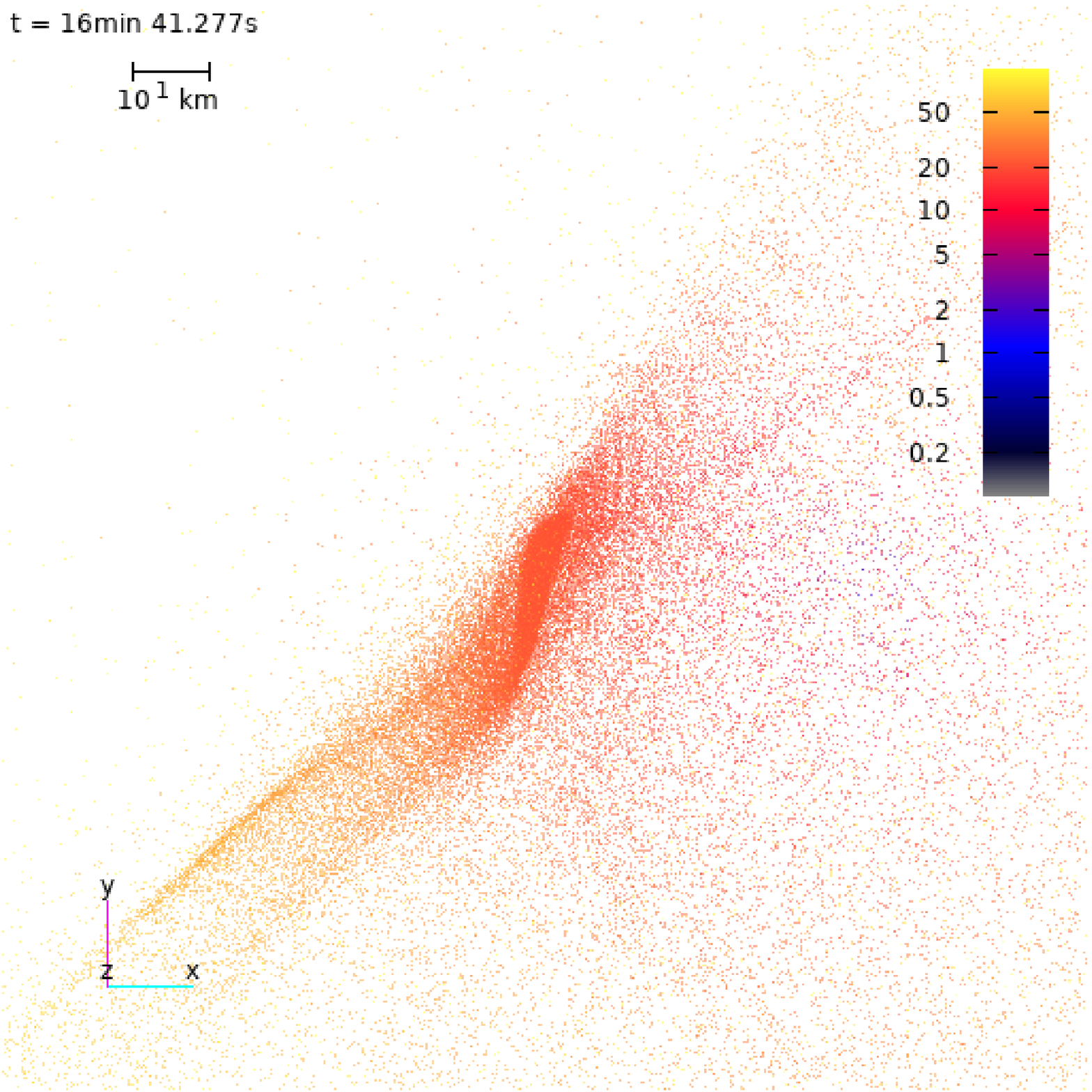} &
\includegraphics[width=\tmpdim]{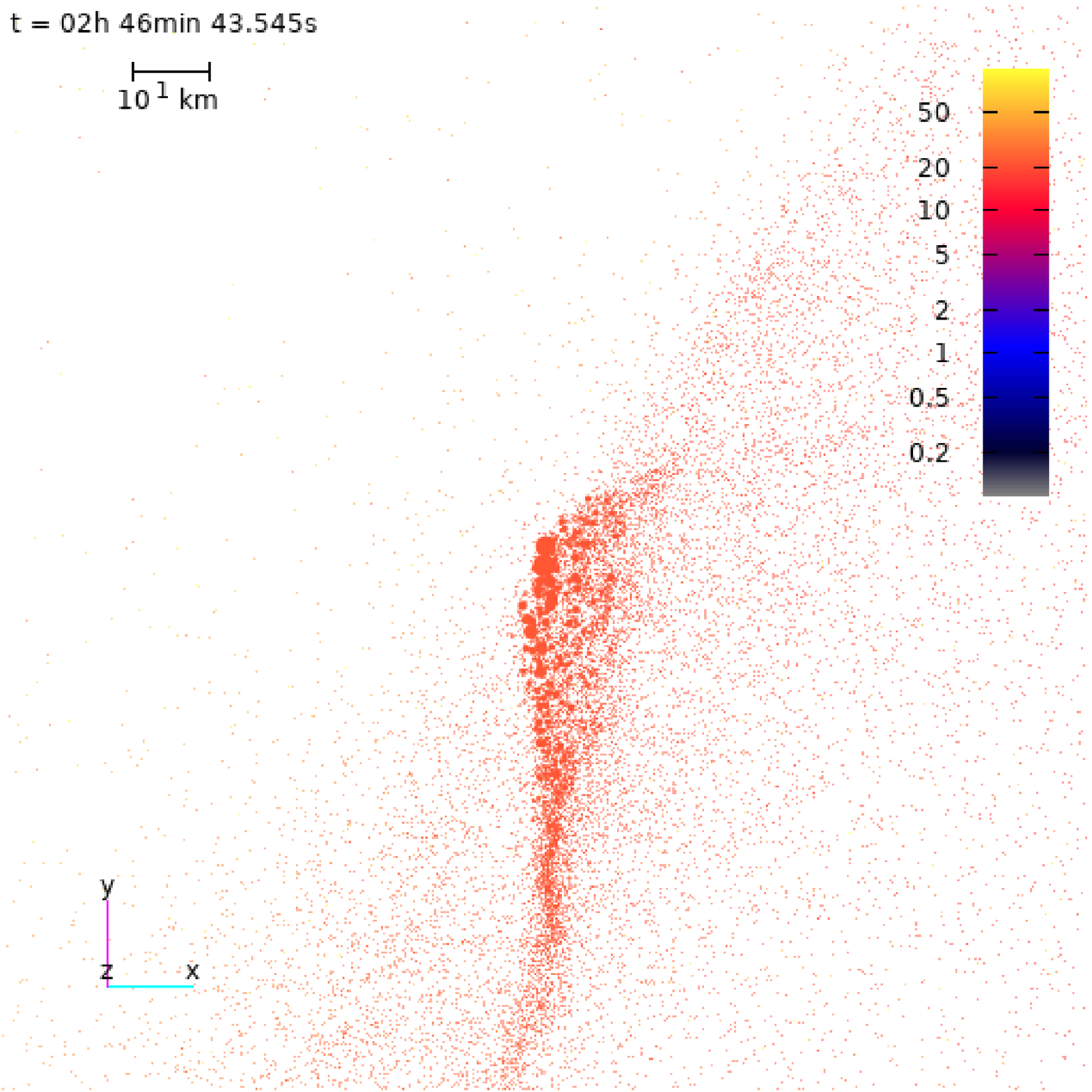} &
\includegraphics[width=\tmpdim]{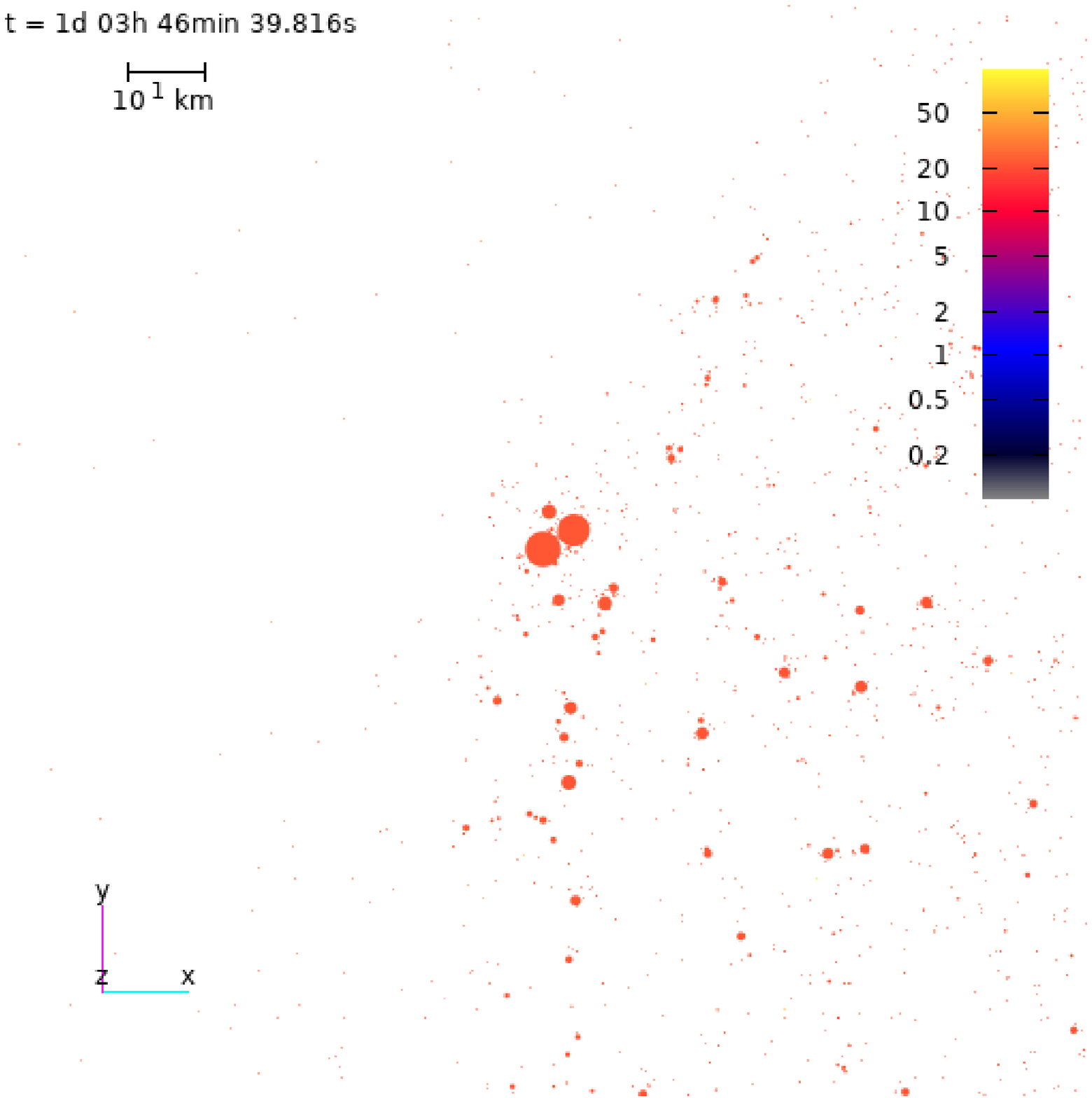} \\
\end{tabular}
\caption{\label{f8}
 Snapshots from an SPH/N-body simulation of a single parent body breakup. The basic parameters
 were a target size $D_{\rm tar} = 9$~km, an impactor size $D_{\rm imp} = 1.3$~km,
 an impact angle $\phi = 30^\circ$ (``near-to-head-on situation''), and an impact
 velocity $v = 5$ km~s$^{-1}$. The fragmentation phase is shown for the epochs
 $t = 0, 10, 100, 10^3$~s ({\em top}, a-d from left to right), and the reaccumulation phase
 for $t = 0, 10^3, 10^4, 10^5$~s (after handoff; {\em bottom}, e-h from left to right).
 The spatial distribution of SPH particles is plotted only within a limited range of the
 coordinate $z \in (-1, 1)$~km to clearly show the interior structure of the parent body,
 its damage, and the clump formation zones. Colors correspond to the velocity $v$ in m~s$^{-1}$
 relative to the target body (scale shown using the bar).
 Individual panels can be described as follows: (a) the initial conditions, (b) high-speed
 ejecta formed at the impact site, (c) formation of a cavity with low relative speeds,
 (d) deformation of the target, (e) handoff phase, (f) streams of high speed,
 individual particles escaping from the system, (g) ongoing reaccumulation,
 and (h) formation of an unbound, nearly equal-size pair accompanied by the subkilometer fragments.
 (Animation is available at \url{https://sirrah.troja.mff.cuni.cz/~mira/hobson/hobson.html})}
\end{figure*}
\begin{figure*}
\centering
\newdimen\tmpdim\tmpdim=4.3cm
\begin{tabular}{cc@{\kern.05cm}c@{\kern.05cm}c@{\kern.05cm}c}
\raise1.0cm\hbox{\rotatebox{90}{fragmentation}} &
\includegraphics[width=\tmpdim]{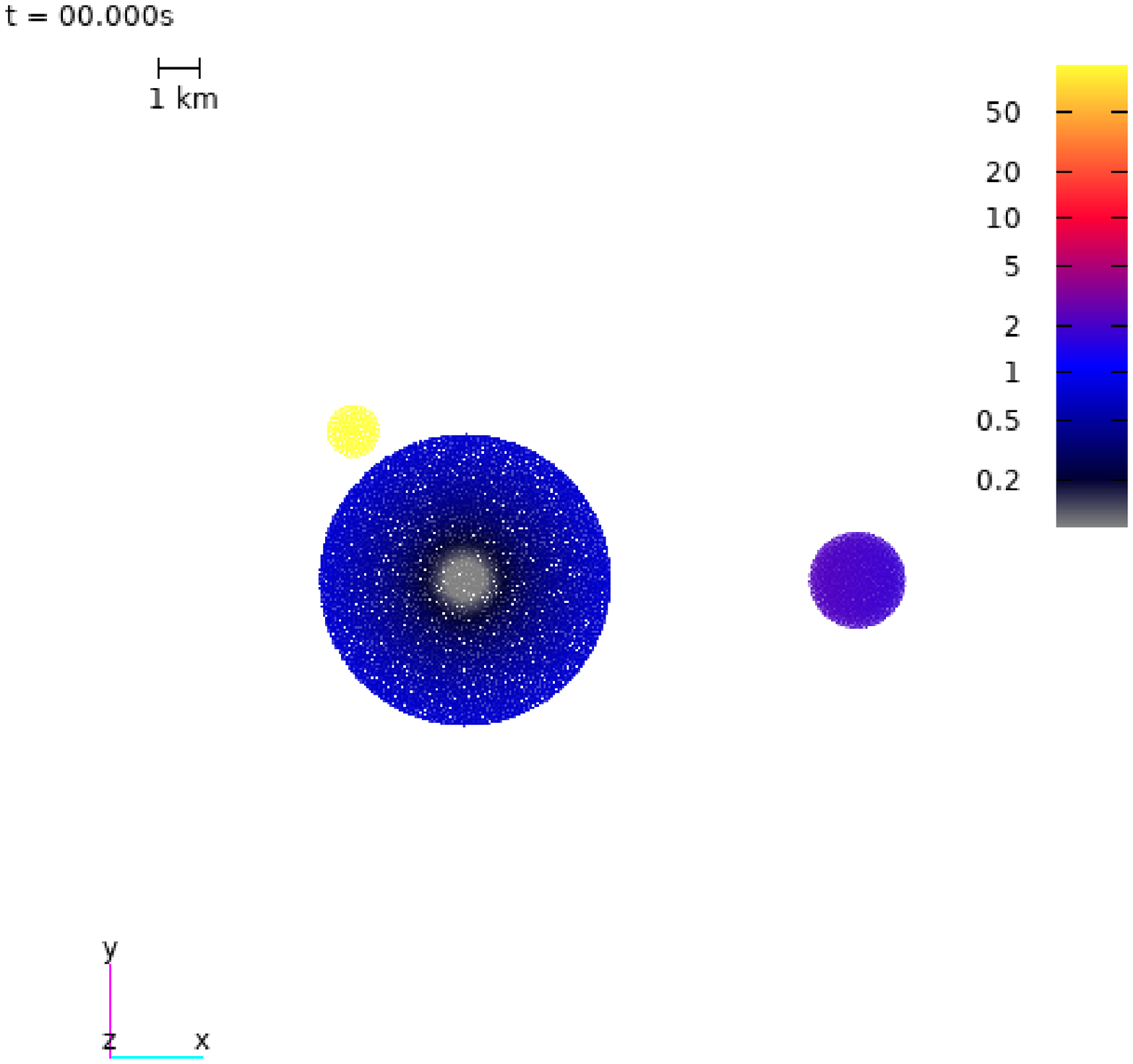} &
\includegraphics[width=\tmpdim]{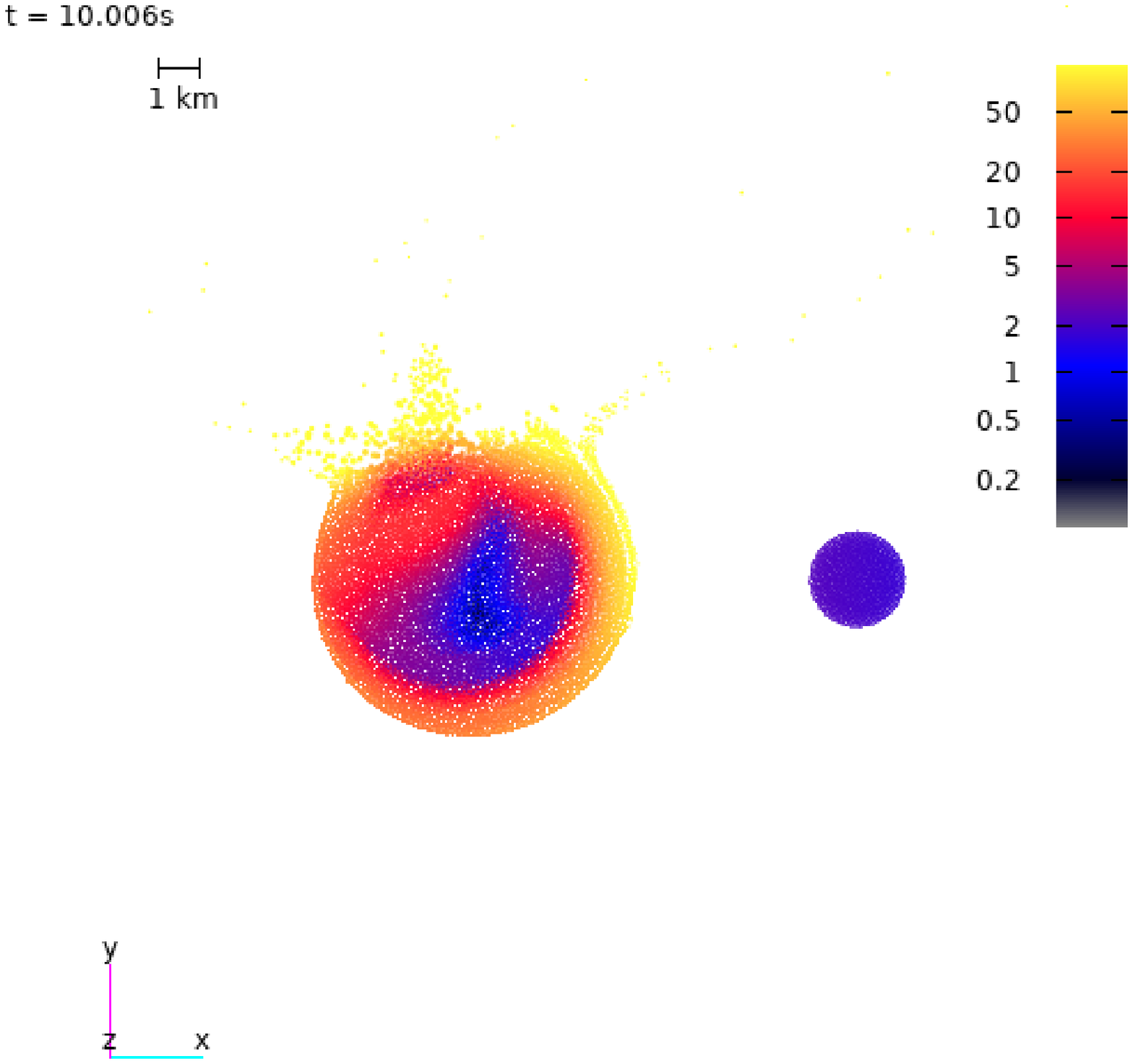} &
\includegraphics[width=\tmpdim]{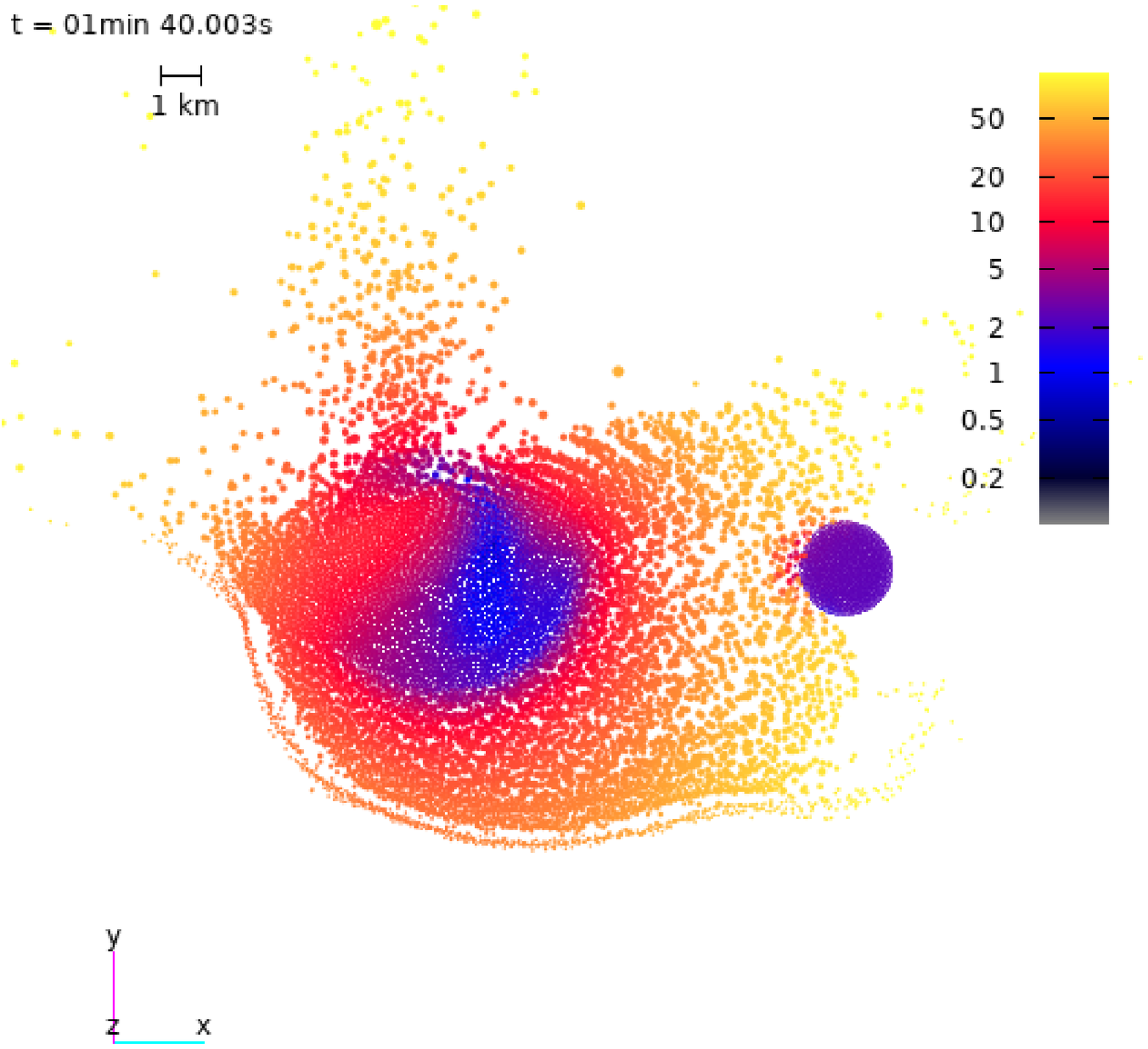} &
\includegraphics[width=\tmpdim]{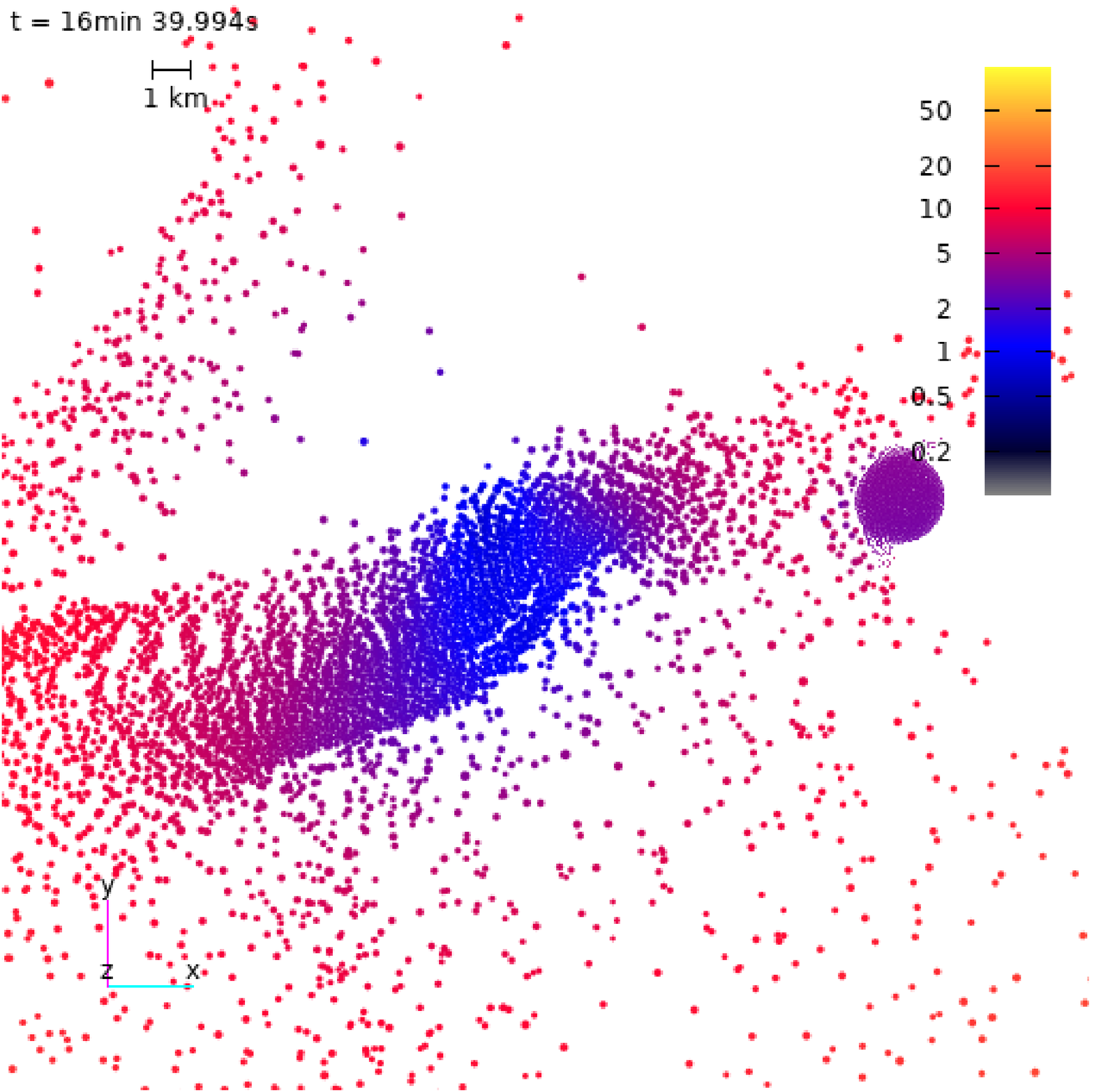} \\[-0.07cm]
\raise0.9cm\hbox{\rotatebox{90}{reaccumulation}} &
\includegraphics[width=\tmpdim]{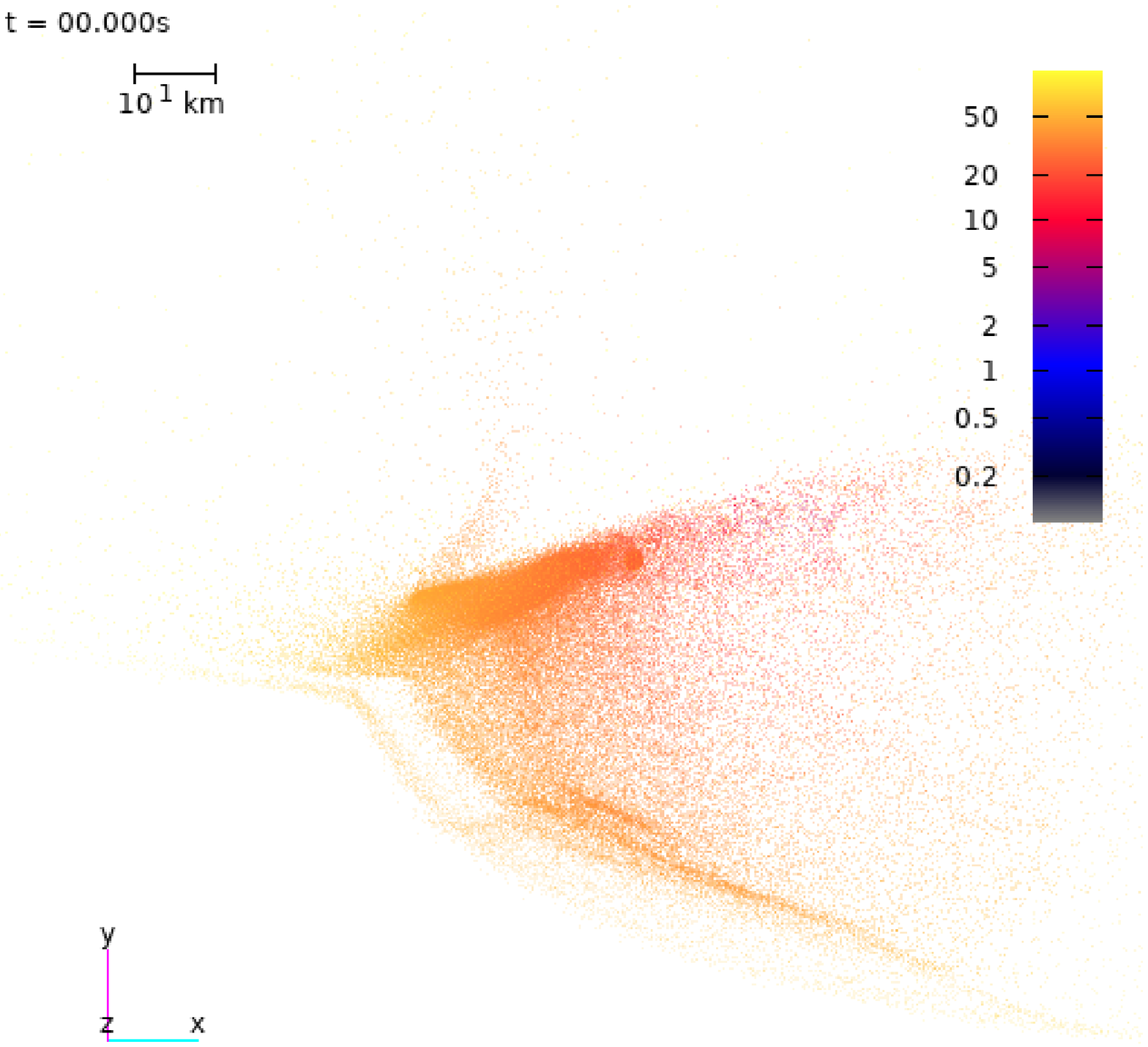} &
\includegraphics[width=\tmpdim]{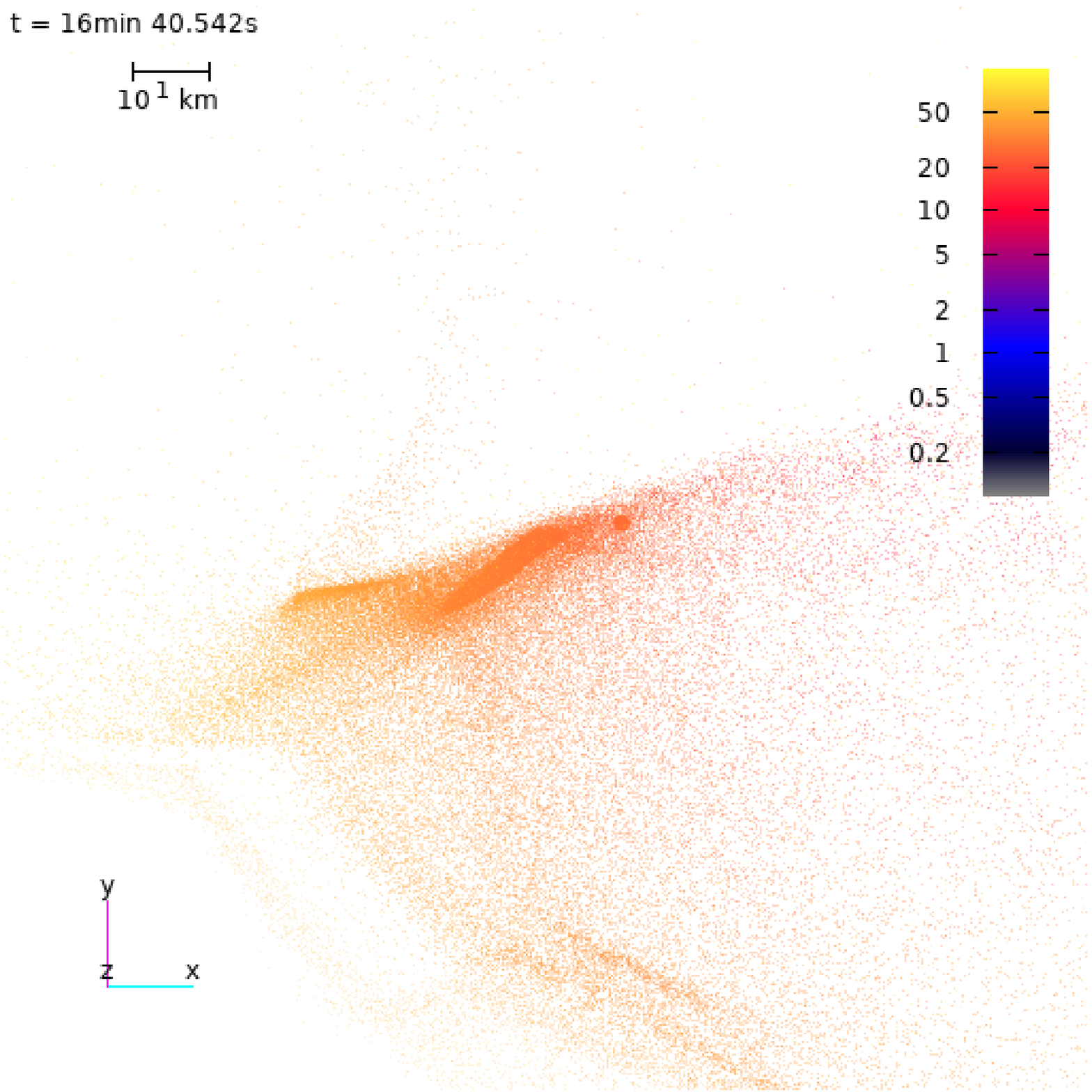} &
\includegraphics[width=\tmpdim]{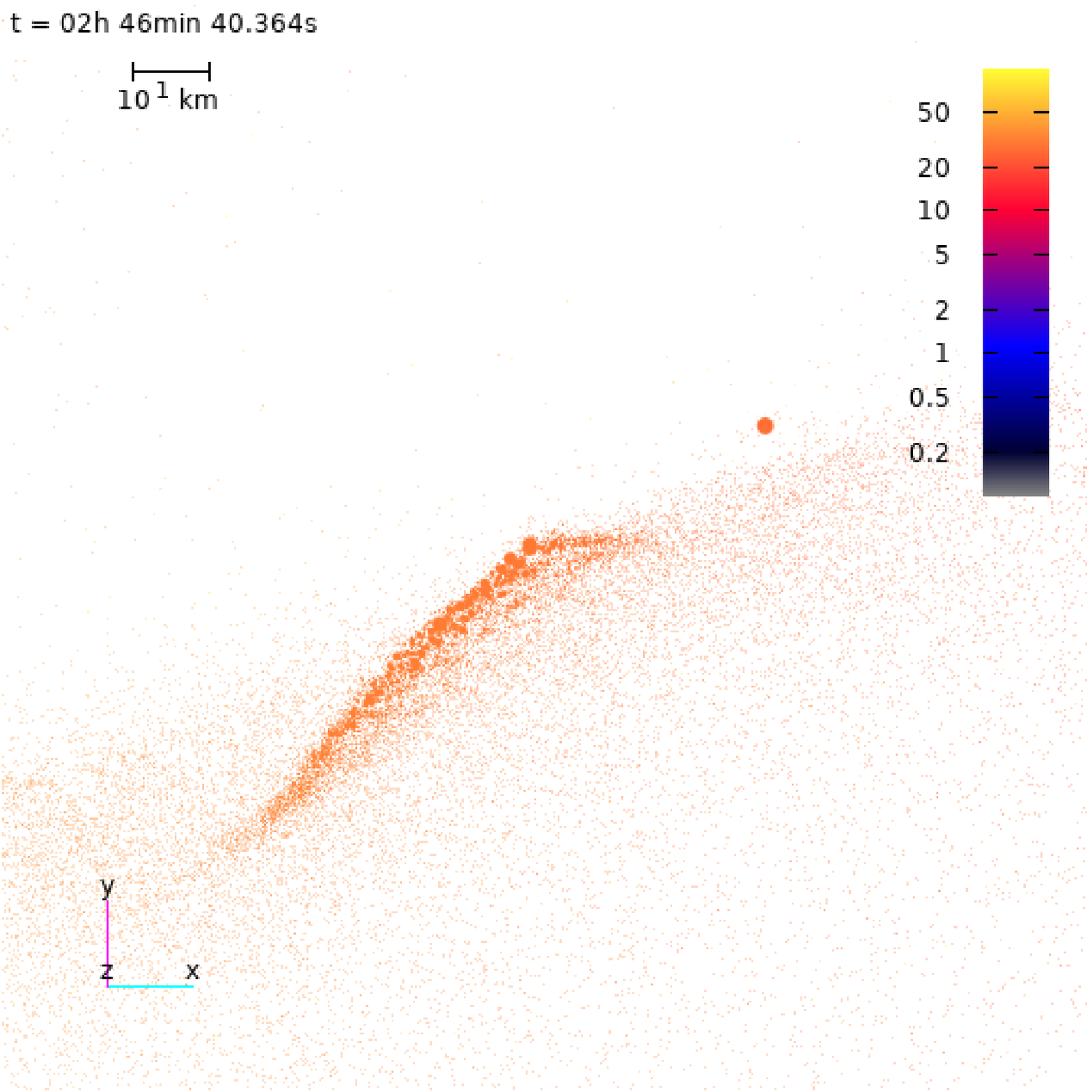} &
\includegraphics[width=\tmpdim]{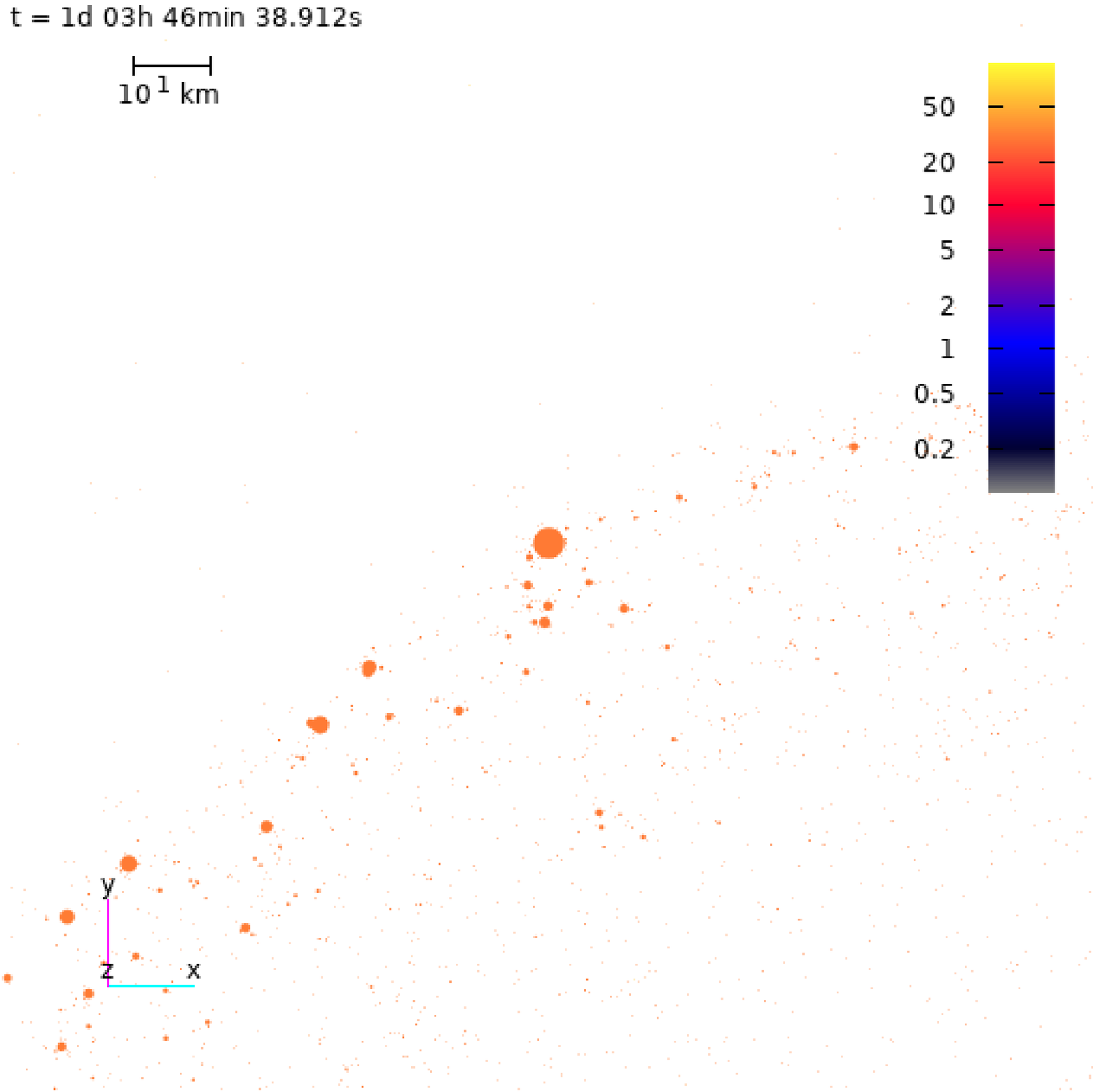} \\
\end{tabular}
\caption{\label{f9}
 Same as Figure~\ref{f8}, but now for a binary parent object. The basic parameters
 were a primary size $D_1 = D_{\rm tar} = 7.5$~km, a secondary size $D_2 = 2.5$~km, an impactor
 size $D_{\rm imp} = 1.4$~km, an impact angle $\phi = 60^\circ$, an impact velocity $v = 5$ km~s$^{-1}$,
 and a separation $r=10$~km. Individual panels can be described as follows:
 (a) the initial conditions with the binary architecture, (b) because of the near-to-grazing
 geometry of the impact, high-speed ejecta emanate from the impact site and
 surrounding surface zone, (c) ejecta reaching the secondary component in the binary,
 (d) the primary is preserved as the largest remnant in the family, (e) handoff phase
 with the preserved secondary component of the binary, (f) onset of reaccumulation
 of other fragments, (g) the secondary escaping from the system, (h) the reaccumulated
 primary (the secondary off-scale, not shown), which together form a distant pair,
 accompanied by smaller fragments. (Animation is available at
 \url{https://sirrah.troja.mff.cuni.cz/~mira/hobson/hobson.html})} 
\end{figure*}

We used the {\tt Opensph} code \citep{setal2019,sevecek2019} for all SPH and N-body
simulations presented below. We substantially improved several aspects of our
previous models or added entirely new features \citep[compare with][]{setal2017}. In particular,
(i) the self-gravity effects were included already in the fragmentation phase,
(ii) and this allowed us to prolong it up to $1000$~s, until the fragments are well separated,
(iii) we implemented a binary architecture of the target body, if needed,  with
 the primary and the secondary components rotating synchronously,
(iv) we abandoned perfect merging in the reaccumulation phase to avoid supercritical rotators,
(v)  we suppressed the merging efficiency to create similar spatial structures as in full SPH runs
(see also \v{S}eve\v{c}ek 2021), and
(vi) we implemented stochasticity testing by performing several simulations with almost the same
 initial conditions. Appendix~B briefly summarizes the setup and parameters used in our
 simulations.

After trial and error, we found two possible solutions that match the SFD properties
of the observed Hobson family (Fig.~\ref{f3}): (i) either a single parent asteroid
breakup, which under special circumstances, results in a similarly sized pair of largest
fragments, or (ii) a binary parent asteroid, with a breakup of the primary and the
secondary (satellite) preserved as a pair component (the special situation of a contact
binary is briefly discussed in Appendix~C). Here we present the most
characteristic examples of both, but we do not intend to accomplish a
detailed scan of the vast parameter space of these simulations. This effort is
postponed to future work.
\smallskip

\noindent{\it Single parent body example. } 
In the case of a single parent body model, we assumed a target size $D_{\rm tar} = 9$~km,
an impactor size $D_{\rm imp} = 1.3$~km, an impact angle $\phi = 30^\circ$, and an impact
velocity $v = 5$ km~s$^{-1}$. The resulting SFD of the synthetic family is shown and
compared with the Hobson-family data in Fig.~\ref{f7} (top panel). While still 
it slightly overestimates the population of small fragments, the simulation is a fairly acceptable
match to the data. Most notably, it provides a pair of $\simeq 2.5-2.7$~km largest remnants,
as seen in the Hobson family. We verified that the two largest remnants are not gravitationally
bound and slowly diverge from each other. According to our tests, the resulting SFD
shown in Fig.~\ref{f7} requires fine-tuning of impact parameters. This is because the
transient pair of the largest remnants most
often merges into a single largest remnant accompanied by a suite of small fragments
\citep[very much like in the results shown by][]{setal2017,setal2019}. More insights into the
underlying mechanism are provided by the sequence of plots showing the spatial distribution
of SPH particles during the fragmentation phase (Fig.~\ref{f8}, top panels). The low
impact angle corresponds to a near-to-head-on collision, fine-tuned to break the parent
asteroid into two similarly sized fragments that diverge from each other just fast enough
to prevent their reaccumulation (Fig.~\ref{f8}, bottom panels). We are currently not
able to fully quantify the statistical likelihood of these special conditions of a breakup. 
This task would require an intense parameter space analysis. This is left for future
work. We needed several dozen trials, but unsuccessful simulations to
finally reach the solution described above. Thus the likelihood is at a few percent at
most, but may be even lower.%
\footnote{The role of the parent body shape is one of the factors that have not been analyzed
 so far. We assumed a spherical shape, but possibly an impact onto a highly elongated body
 (such as the near-Earth asteroid 1620 Geographos) may more easily result in the formation of two
 nearly equal-size largest remnants.}
\smallskip

\noindent{\it Binary parent body example. } 
For a binary model, we assumed the following parameters: a primary size $D_1 = D_{\rm tar} =
7.5$~km, a secondary size $D_2 = 2.5$~km, an impactor size $D_{\rm imp} = 1.4$~km, an orbital and spin
rate(s) $\omega = 17.55$~d$^{-1}$, an impact angle $\phi = 60^\circ$, an impact velocity
$v = 5$ km~s$^{-1}$, a binary separation $r = 10$~km, and an orbital velocity $v_{\rm orb} =
2$ m~s$^{-1}$. We intentionally considered a more compact proto-binary than suggested
in Sec.~\ref{sfd} (in particular, the secondary rotation period would only be $\simeq 9$~hr),
with the goal to test collisional fate of the secondary in the most severe regime.
The resulting synthetic SFD is now shown in the bottom panel of Fig.~\ref{f7}.
The second largest remnant in the synthetic population is the fully preserved, unbound
secondary of the parent binary system. The remaining populations of bodies consist of the
shattered primary, including the largest remnant. Again, the small number of tests we
did at this stage does not permit a very detailed tuning in comparison with the
data. For instance, in our simulation the largest remnant is slightly smaller than needed,
at the expense of a more significant population of smaller fragments. Fine-tuning of
the impact parameters would certainly allow an even better comparison between data and
model. An intense effort in this direction would, however, only be successful when the
observed population of Hobson members is corrected by the observational incompleteness.
Here we only mention that we tried several simulations by slightly changing the impact
velocity $v = 4.96$ to $5$ km~s$^{-1}$ and the impact angle $\phi = 60$ to $60.5^\circ$.
Most often, the results were similar to what is shown in Fig.~\ref{f7}, but sometimes
the primary was shattered too much (such that the largest fragment became even smaller
than the secondary). We note that the results are insensitive to the initial
separation $r$ of the parent binary system, in particular, wider binaries with $r=20-30$~km 
would provide still the same results. The secondary would become more easily
unbounded and would certainly be preserved intact.
Even in our compact variant with $r=10$~km, the interior of the secondary never experiences
damage or heating. In this respect, it is a singular member in the future family,
but it is not clear what observation would allow us to determine this property.
\smallskip
\begin{figure}[t]
 \begin{center} 
 \includegraphics[width=0.49\textwidth]{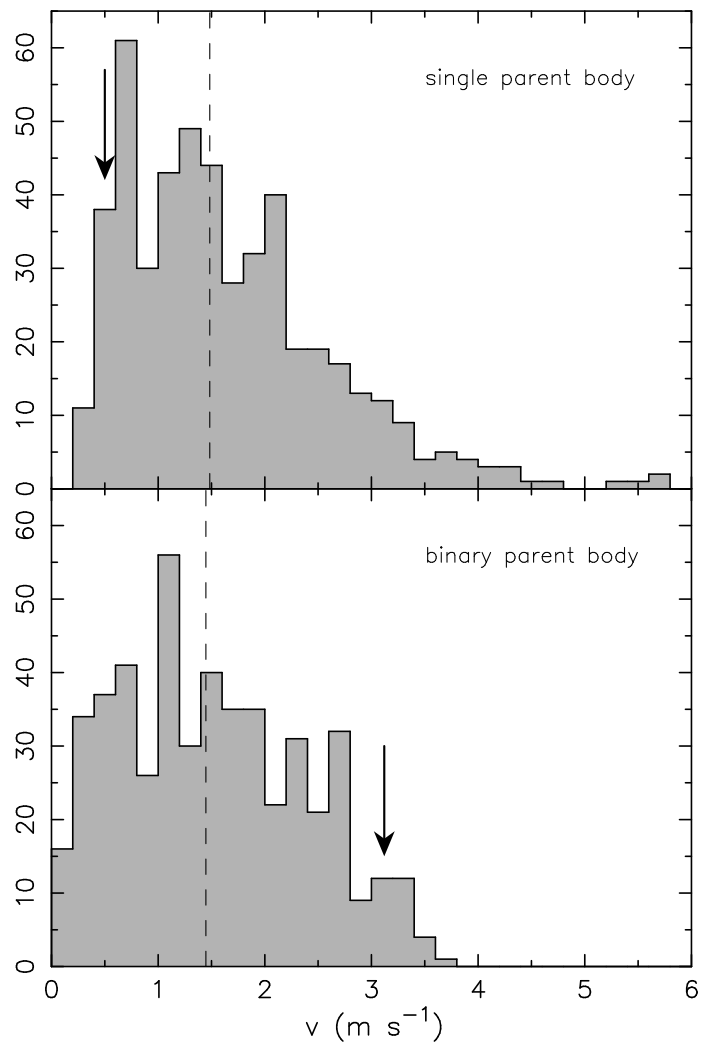}
 \end{center}
 \caption{\label{f10}
  Distribution of the relative velocity of the $500$ largest fragments (sizes approximately
  $\geq 300$~m; Fig.~\ref{f7}) with respect to the largest remnant in our synthetic
  realizations of the Hobson family: (i) the case of a single parent body ({\em top}),
  and (ii) the case of binary parent system ({\em bottom}). The number of fragments
  in $0.2$ m~s$^{-1}$ wide bins on the ordinate. The median value $\simeq 1.5$ m~s$^{-1}$
  is shown by the dashed vertical line. The arrow indicates the velocity of the second largest
  remnant (57738) 2001~UZ160.}
\end{figure}

\noindent{\it Dispersal velocity of the largest fragments. }In both models above, we
also determined the final velocity dispersal with respect to the largest remnant (i.e., 
18777~Hobson). Panels e) and f) in the bottom rows of Figs.~\ref{f8} and \ref{f9} would 
suggest at first glance that the geometries of the velocity fields are very different. This is 
indeed true, but it concerns mostly the fast-escaping, resolution-level single particles in 
our simulations. The properties of the velocity
fields for multi-particle clumps, representing real fragments, are more similar to each
other, especially when they are restricted to the set of the first few hundred largest fragments
(this group overlaps with the sizes of the observed Hobson members; Fig.~\ref{f7}).
For the sake of simplicity, we did not consider the directions of the relative velocity vectors
with respect to the largest fragment, but only their magnitude. Figure~\ref{f10} shows these
results for both models, single asteroid and binary system parents. In both of them
the velocities of the largest fragments (here with sizes $\geq 300$~m) are in the range $0$ to
$4$ m~s$^{-1}$, with a median of about $1.5$ m~s$^{-1}$. A notable difference consists of the
relative velocity of the second largest remnant, namely (57738) 2001~UZ160 (indicated by the
arrow): this value is very low in the first model ($\simeq 0.5$ m~s$^{-1}$), but much higher 
in the second model ($\simeq 3.2$ m~s$^{-1}$; see also Fig.~\ref{f9}, bottom panels).
However, all these velocities remain very low and satisfy the upper limits set by
the backward convergence tests in Sec.~\ref{conv} very well (although we recall a hint of the
57738 2001~UZ160 offset from other small members in the Hobson family reported at the
end of that section). The escape velocity from the modeled
parent systems is $5-6$ m~s$^{-1}$, which is slightly higher than the characteristic velocity
dispersal of observable subkilometer fragments. The backward numerical integrations over
hundreds of thousands of years are not deterministic enough to trace the tiny model
difference in the relative velocity of the two largest remnants in the Hobson family
when all sources of uncertainty are taken into account (Sec.~\ref{pair}). Unfortunately,
the highest fragment velocities thus cannot distinguish between the two models for the parent
system of the Hobson family. More detailed analyses are needed in the future.

\section{Discussion and conclusions} \label{disc}
It is interesting to compare our results with the much more detailed information about
the Karin family, an archetype in the category of young families. Its quite larger known
population of fragments allowed \citet{karin2006} a detailed comparison of the family
structure and size distribution with model predictions. Here we focus primarily on the conclusions
related to the size distribution and velocity dispersal of the observed fragments.

\citet{karin2006} found that the Karin family was formed by an impact of $\simeq 5.8$~km
projectile onto an  $\simeq 33$~km parent body with $\simeq 6-7$ km~s$^{-1}$ speed and $\simeq
45^\circ$ impact angle (average in the main belt). The largest created remnant, (832)~Karin,
is about $17$~km, and the size distribution of the Karin fragments then follows from the
second largest fragment of $\simeq 5.5$~km with a steep cumulative power law of $\simeq -5.3$
exponent (which translates into a cumulative absolute magnitude exponent $\gamma_{\rm Karin}
\simeq 1.06$, comparable with the Hobson population; Sec.~\ref{sfd}). Modeling thus indicates
that Karin resulted from
a rather energetic collision with $\simeq (0.1-0.15)$ mass ratio between the largest
remaining fragment and the parent body. It is interesting to note that the size distribution
of fragments in the Karin and Hobson families are in many respects similar, but also have
intriguing differences: (i) given the plausible observational incompleteness of the small Hobson
members, it is possible that the power-law exponents of the size distribution section
starting from the second and third largest fragments are similar (or even slightly steeper
for Hobson), and (ii) the gap of $\simeq (2-2.5)$~magnitude between the largest fragment(s) and
the continuum section is also comparable in the two families. The Karin model thus fits our 
second, binary scenario for the Hobson family formation very well because the fragmentation
properties of its primary compare well to what we see for Karin. The main
difference clearly consists of the fact that the parent body of the Karin family was not binary,
and thus the Karin family misses the second largest remnant of a comparable size to (832) Karin.
As to the ejection velocities $v_{\rm ej}$, \citet{karin2006} found that $D_{\rm f}\simeq 3$~km
fragments (typical for the steep leg in the size distribution) have mean barycentric ejection
speeds of $\simeq 12$ m~s$^{-1}$ with the fastest (and smallest) launched at 
$\simeq 30$ m~s$^{-1}$. These values are comparable to or only slightly higher than the 
escape speed from the parent asteroid, namely $\simeq 20$ m~s$^{-1}$. This is again a
similar ratio as in the simulations of the Hobson family.

Our findings indicate that many of the Karin family results may also hold for the Hobson
family. Overall, the similarity of the size distributions has been mentioned above. 
The main characteristics of the dispersal velocity field of the observable fragments 
may also be similar if scaled by the estimated escape velocity from the parent body. 
The results in the Hobson case are clearly much less accurate at this moment. They will hopefully
significantly improve when the next decade of sky surveys will allow us to discover many
more small fragments and even provide their physical characterization.

The significance of the Karin to Hobson comparison is especially highlighted by observing
the difference in their estimated parent body size: (i)
$\simeq 33$~km for the Karin family, and (ii) $\simeq 7-9$~km for the Hobson family. The quite
larger size in the Karin case helps gravity to hold the fragments and moderate their typical
dispersal velocities. Probing with Hobson a regime of much smaller parent objects is important 
because the gravity becomes much weaker in this case. We consider, for instance, that the specific energy
for the disruption $Q^\star_{\rm D}$ is more than an order of magnitude lower for the
Hobson parent body \citep[e.g.,][]{ba1999,betal2015}.
\smallskip

\noindent{\it Likelihood of the Hobson family formation from a parent binary.}
We next verified that our proposed formation of the Hobson family from the collisional
disruption of a primary component of the main belt binary is justifiable
in a statistical sense (plausibly assuming the same statistical properties of small
binaries in its inner and middle parts). We first consulted the results shown in Fig.~15
of \citet{betal2005},
where a characteristic timescale for the disruption of a main belt asteroid of a given size
was determined. Their simulation considered the
main belt as a whole and clearly did not resolve solitary and binary objects.
The information is provided as a function of size only, which we associate with
the $\simeq 8$~km size of the primary of our proposed binary object for the Hobson
family. From this, we obtain a characteristic disruption timescale of
$\simeq (30-50)$~kyr. Taking into account that only every sixth to seventh is
a binary \citep{petal2016}, the timescale becomes $\simeq (180-350)$~kyr. From
this point of view, the $300$ to $400$~kyr age of the Hobson
family appears very plausible. The recent origin of the Hobson family from a binary
does not need to be considered a statistical fluke.

As discussed in Sec. 3, an alternative possibility to our model of the binary parent
body is a more traditional assumption of a single parent body. Because the required 
parent size is similar to the primary of the binary system discussed
above, the \citet{betal2005} collisional model provides many such breakups in
the past $500$~kyr, for example. However, the formation of a pair of nearly equal-size
largest remnants in the family needs special impact conditions (maybe
with a probability of only 1\%). This issue needs to be studied in more
detail in future work.
\smallskip

\noindent{\it Likelihood of a hypothetical Hobson-2001~UZ160 binary split.}
As discussed in Sec.~\ref{sfd_mod}, an alternative possibility to our model of
a binary parent body is the more traditional assumption of a single parent body
(requiring then special impact parameters). A pair of nearly
equal-size largest remnants may be formed, but the question is whether
these two objects more likely separate immediately into two asteroids on different
heliocentric orbits, or create a bound binary that subsequently underwent instability.
Here we show that the latter case is rather unlikely, and the immediate
separation of (18777)~Hobson and (57738) 2001~UZ160 is the preferred case.

First, we considered a scenario in which a putative binary disrupted due to a subcatastrophic
impact onto one of the components. Assuming conservatively the orbital velocity of
$\simeq 0.5$ m~s$^{-1}$, we used the formulas given in \citet{netal2011} and \citet{netal2019} to
estimate that the required impactor size is $\simeq 150$~m. Already this
information appears to contradict the assumption of a nondisruptive event. This
is because the critical impact specific energy for a $\simeq 2.5$~km target is
low: $Q_{\rm D}^\star \simeq 700$ J~kg$^{-1}$ \citep[e.g.,][]{betal2005,betal2020}.
Therefore conventionally, when we assume a characteristic
impact velocity of $\simeq 5$ km~s$^{-1}$, an impactor of $\simeq 100$~m
would produce a catastrophic disruption of either component in the hypothetical
binary. Even if we were generous and assumed a loose binary system prone to separation
whose critical impact may be as small as $\simeq 50$~m, the idea would
not hold. This is because the probability of such an impact in $\simeq 400$~kyr
is very low. In order to show how large, we took the characteristic intrinsic
collisional probability $P_{\rm i}\simeq 2.9\times 10^{-18}$ km$^{-2}$~yr$^{-1}$ of
the objects in the main belt and considered that there are $N_{\rm imp}\simeq 10^8$ such impactors
\citep[e.g.,][]{betal2005,betal2020}. In $T\simeq 400$~kyr (the
estimated age of the Hobson family), we therefore expect $\simeq N_{\rm imp} P_{\rm i} R^2 T
\simeq 2\times 10^{-4}$ of such events to happen ($R\simeq 1.25$~km is the estimated
characteristic size in the Hobson-2001~UZ160 binary). There is simply not enough
time since the formation of the Hobson family for the hypothetical binary
to split collisionally.

Another possibility is that the hypothetical Hobson-2001~UZ160 binary split dynamically.
The most likely
candidate process would be the binary Yarkovsky-O'Keefe-Radzievskii-Paddack (BYORP)
radiative effect \citep[e.g.,][]{cb2005}. Here again the likelihood is very low, even taking
the most aggressive scenario \citep[e.g., assuming a permanent spin-orbital
synchronous state without interruptions; see, e.g.,][]{cn2010}. In this case, the BYORP
instability timescale would simply be the typical YORP timescale of a $\simeq 2.5$~km
size asteroid at the heliocentric distance of the Hobson family, typically $\simeq 10$~Myr
\citep[e.g.,][]{cb2005}. This is again much longer than the estimated age of the
Hobson family.
\smallskip

\noindent{\it Future prospects.} The Hobson family will continue to be an interesting 
example of very young asteroid families. The experience from the past decade shows that
its known population may easily double in the next few years, especially if powerful surveys
such as the Vera C. Rubin Observatory will reach their expected operations. This will
allow us not only to improve our analysis in this paper, but perhaps tackle other issues.
Refining solution of the Hobson family age, guided by the synthetic model of the family
formation, may be one of the most interesting projects. Eventually, this might
help distinguishing among the two models of the parent system.

\begin{acknowledgements}
 We thank the referee whose comments and suggestions helped to improve the submitted
 version of this paper. 
 This research was supported by the Czech Science Foundation (DV and MB; grant~21-11058S).
 BN acknowledges the support of the Ministry of Education, Science and Technological
 Development of the Republic of Serbia, contract No.~451-03-68/2020-14/200104.

\end{acknowledgements}


\begin{appendix}

\section{Hobson family. Membership and proper elements}
In this section, we provide a complete list of $45$ Hobson family members determined
in Sec.~\ref{iden} and we represent the family using the traditional set of proper
orbital elements: semimajor axis $a_{\rm P}$, eccentricity $e_{\rm P}$, and sine of
the inclination $\sin I_{\rm P}$.  We use synthetic elements, determined by
numerical integration \citep[e.g.,][]{km2000,km2003}. As mentioned in the appendix of
\citet{vetal2021}, we cannot use data from the standard world storehouses of proper
orbital elements, such as {\tt AstDyS}%
\footnote{\url{https://newton.spacedys.com/astdys/}}
or {\tt AFP}%
\footnote{\url{http://asteroids.matf.bg.ac.rs/fam/}}.
This is because many of small Hobson members have been discovered only
recently, and these databases have not been updated yet, or have rejected some orbits
with too few observations at this moment. As a result, we computed the proper elements
ourselves, using the methods described in the appendix of \citet{vetal2021}.
Here we outline the principal steps.
\begin{figure}[t]
 \begin{center} 
 \includegraphics[width=0.49\textwidth]{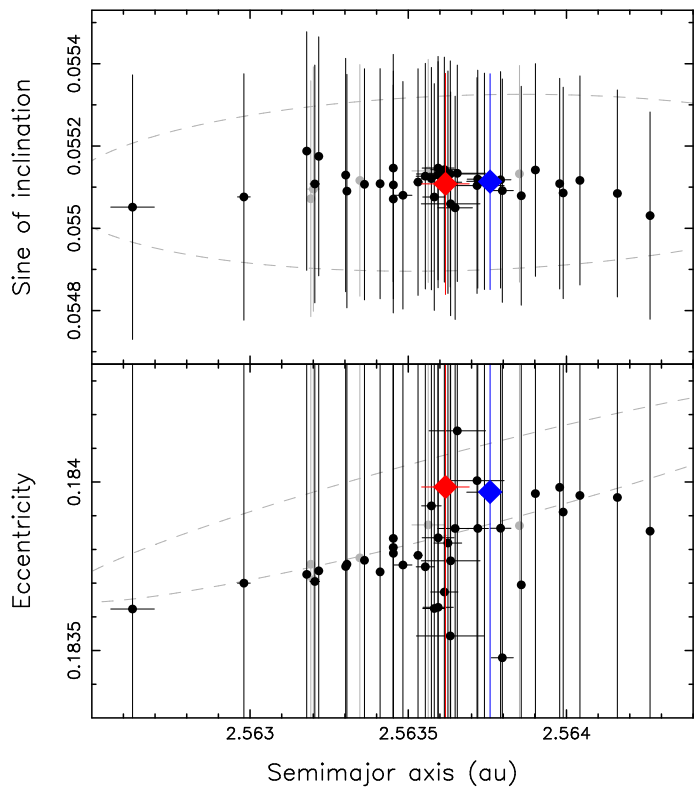}
 \end{center}
 \caption{\label{fa1}
  Hobson family represented by two possible plane projections of the proper orbital
  elements: (i) semimajor axis $a_{\rm P}$ vs. sine of the inclination $\sin I_{\rm P}$ in the
  top panel, and (ii) semimajor axis $a_{\rm P}$ vs. eccentricity $e_{\rm P}$ in the
  bottom panel. The largest fragments (18777)~Hobson (red) and (57738) 2001~UZ160
  (blue) are shown by diamonds, and smaller family members are shown by filled circles
  (gray are the five single-opposition orbits). Vertical and horizontal bars are formal
  uncertainty values of the respective proper element from Table~\ref{tab_prop}.
  The uncertainty in $e_{\rm P}$ is large due to perturbing effect of the $g+g_5-2g_6$
  secular resonance. The gray dashed ellipses indicate proper element zones in which
  fragments may land if they are ejected isotropically from the barycenter of the family with
  $4$~m~s$^{-1}$ velocity (assuming the true anomaly $f=50^\circ$ and the argument of perihelion
  such that $\omega + f = 0^\circ$). The tilt in the bottom panel may be
  due to the perturbing effect of the $g+g_5-2g_6$ resonance, however. The particular dispersion
  of the family members in $e_{\rm P}$ at $a_{\rm P}\simeq 2.5636$~au, with the noticeable
  uncertainty of the $a_{\rm P}$ values, is due to the perturbation by the 9J-8S-2 three-body
  mean-motion resonance.}
\end{figure}
\begin{table*}[ht]
\caption{\label{tab_prop}
 Proper orbital elements and their formal uncertainty of the Hobson family members.}
\centering
\begin{tabular}{rlccccccc}
\hline \hline
 \multicolumn{2}{c}{Asteroid} & \rule{0pt}{2ex} $a_{\rm P}$ & $\delta a_{\rm P}$ &
  $e_{\rm P}$ & $\delta e_{\rm P}$ & $\sin I_{\rm P}$ & $\delta \sin I_{\rm P}$ &
  $H$ \\
 & & [au] & [au] & & & & & [mag] \\
\hline
\rule{0pt}{3ex}
 18777 & Hobson     & 2.5636173 & 0.0000751 & 0.18399 & 0.01146 & 0.055108 & 0.000269 & 15.16 \\    
 57738 & 2001~UZ160 & 2.5637584 & 0.0000193 & 0.18397 & 0.01149 & 0.055113 & 0.000262 & 15.41 \\    
363118 & 2001~NH14  & 2.5635306 & 0.0000139 & 0.18378 & 0.01147 & 0.055112 & 0.000275 & 17.35 \\    
381414 & 2008~JK37  & 2.5639014 & 0.0000066 & 0.18397 & 0.01154 & 0.055142 & 0.000259 & 17.69 \\    
436620 & 2011~LF12  & 2.5637171 & 0.0000853 & 0.18400 & 0.01148 & 0.055104 & 0.000262 & 17.33 \\    
450571 & 2006~JH35  & 2.5634527 & 0.0000032 & 0.18379 & 0.01144 & 0.055105 & 0.000277 & 17.40 \\    
465404 & 2008~HQ46  & 2.5634525 & 0.0000032 & 0.18381 & 0.01143 & 0.055071 & 0.000277 & 17.73 \\    
520394 & 2014~JJ10  & 2.5636325 & 0.0001072 & 0.18354 & 0.01155 & 0.055132 & 0.000275 & 17.90 \\    
537249 & 2015~HM190 & 2.5635535 & 0.0000289 & 0.18375 & 0.01147 & 0.055127 & 0.000274 & 17.61 \\    
548822 & 2010~VG231 & 2.5639892 & 0.0000080 & 0.18391 & 0.01156 & 0.055086 & 0.000257 & 17.90 \\    
557505 & 2014~UB262 & 2.5632046 & 0.0000150 & 0.18370 & 0.01138 & 0.055108 & 0.000289 & 18.38 \\    
       & 2007~EH116 & 2.5637914 & 0.0000327 & 0.18386 & 0.01156 & 0.055118 & 0.000263 & 17.60 \\    
       & 2007~HC54  & 2.5636337 & 0.0000929 & 0.18377 & 0.01149 & 0.055060 & 0.000272 & 17.10 \\    
       & 2008~WV149 & 2.5633020 & 0.0000070 & 0.18375 & 0.01141 & 0.055129 & 0.000284 & 17.80 \\    
       & 2010~GN203 & 2.5635940 & 0.0000479 & 0.18363 & 0.01151 & 0.055130 & 0.000275 & 17.90 \\    
       & 2011~SU302 & 2.5634525 & 0.0000034 & 0.18383 & 0.01144 & 0.055146 & 0.000276 & 18.40 \\    
       & 2012~JM71  & 2.5636546 & 0.0000895 & 0.18415 & 0.01147 & 0.055134 & 0.000263 & 18.10 \\    
       & 2012~LN31  & 2.5639782 & 0.0000072 & 0.18398 & 0.01155 & 0.055109 & 0.000256 & 18.10 \\    
       & 2013~MW20  & 2.5637969 & 0.0000349 & 0.18349 & 0.01157 & 0.055092 & 0.000272 & 18.10 \\    
       & 2014~HH103 & 2.5635819 & 0.0000366 & 0.18362 & 0.01150 & 0.055076 & 0.000276 & 17.80 \\    
       & 2014~KY102 & 2.5638569 & 0.0000102 & 0.18369 & 0.01155 & 0.055079 & 0.000266 & 17.90 \\    
       & 2014~NN71  & 2.5637194 & 0.0000336 & 0.18386 & 0.01150 & 0.055120 & 0.000265 & 18.10 \\    
       & 2014~OG277 & 2.5629804 & 0.0000200 & 0.18370 & 0.01130 & 0.055076 & 0.000299 & 18.40 \\    
       & 2014~PJ87  & 2.5631790 & 0.0000053 & 0.18373 & 0.01139 & 0.055188 & 0.000290 & 18.30 \\    
       & 2014~QL520 & 2.5636478 & 0.0000547 & 0.18386 & 0.01146 & 0.055050 & 0.000271 & 18.30 \\    
       & 2015~FV225 & 2.5633059 & 0.0000059 & 0.18376 & 0.01140 & 0.055090 & 0.000284 & 17.60 \\    
       & 2015~HV138 & 2.5637405 & 0.0000559 & 0.18397 & 0.01149 & 0.055114 & 0.000263 & 18.70 \\    
       & 2015~KA91  & 2.5640422 & 0.0000065 & 0.18396 & 0.01157 & 0.055116 & 0.000254 & 17.90 \\    
       & 2015~OP104 & 2.5634829 & 0.0000285 & 0.18375 & 0.01145 & 0.055080 & 0.000277 & 18.00 \\    
       & 2015~PM156 & 2.5641606 & 0.0000060 & 0.18395 & 0.01159 & 0.055085 & 0.000252 & 18.40 \\    
       & 2015~PA184 & 2.5626285 & 0.0000688 & 0.18362 & 0.01118 & 0.055051 & 0.000322 & 19.20 \\    
       & 2015~XL282 & 2.5633615 & 0.0000061 & 0.18377 & 0.01141 & 0.055107 & 0.000281 & 17.60 \\    
       & 2016~GY256 & 2.5632169 & 0.0000047 & 0.18374 & 0.01140 & 0.055175 & 0.000291 & 18.00 \\    
       & 2016~GW276 & 2.5636250 & 0.0000436 & 0.18382 & 0.01148 & 0.055112 & 0.000270 & 18.30 \\    
       & 2017~SQ83  & 2.5635942 & 0.0000506 & 0.18383 & 0.01147 & 0.055146 & 0.000271 & 18.10 \\    
       & 2017~WO47  & 2.5635729 & 0.0000311 & 0.18393 & 0.01145 & 0.055121 & 0.000270 & 17.80 \\    
       & 2019~NP44  & 2.5636140 & 0.0000430 & 0.18367 & 0.01152 & 0.055143 & 0.000274 & 18.90 \\    
       & 2019~PS30  & 2.5634107 & 0.0000059 & 0.18373 & 0.01143 & 0.055108 & 0.000280 & 18.60 \\    
       & 2020~HQ57  & 2.5642638 & 0.0000100 & 0.18385 & 0.01162 & 0.055031 & 0.000252 & 18.50 \\ [2pt]   
       & \e{2014~JH120} & \e{2.5635629} & \e{0.0000515} & \e{0.18387} & \e{0.01144} & \e{0.055140} & \e{0.000271} & \e{18.70} \\    
       & \e{2014~OJ66}  & \e{2.5638510} & \e{0.0000095} & \e{0.18387} & \e{0.01153} & \e{0.055132} & \e{0.000263} & \e{18.60} \\    
       & \e{2020~JM31}  & \e{2.5633468} & \e{0.0000059} & \e{0.18378} & \e{0.01141} & \e{0.055116} & \e{0.000282} & \e{18.50} \\    
       & \e{2020~KP36}  & \e{2.5644773} & \e{0.0000059} & \e{0.18408} & \e{0.01165} & \e{0.055055} & \e{0.000240} & \e{18.90} \\    
       & \e{2020~OY50}  & \e{2.5632000} & \e{0.0000188} & \e{0.18371} & \e{0.01137} & \e{0.055095} & \e{0.000297} & \e{18.60} \\
       & \e{2021~MO5}   & \e{2.5631920} & \e{0.0000114} & \e{0.18376} & \e{0.01137} & \e{0.055072} & \e{0.000287} & \e{19.00} \\ [2pt]
\hline
\end{tabular}
\tablefoot{Hobson family membership as of July 2021. The first column lists
 the asteroid number (if numbered) and identification. The next six columns provide the asteroid
 proper elements $(a_{\rm P},e_{\rm P},\sin I_{\rm P})$ and their formal
 uncertainty $(\delta a_{\rm P},\delta e_{\rm P},\delta \sin I_{\rm P})$ determined
 by the methods described in Appendix~A. The last column gives the absolute magnitude
 $H$ from MPC database. Being a byproduct of orbit determination procedure
 from observations of sky surveys, the listed $H$ values might be uncertain.
 The exception are the largest two members, 18777 and 57738, whose values were determined
 using well-calibrated photometric observation by \citet{petal2018}. As a result, their
 uncertainty is only $0.05$~magnitude.
 Asteroids whose data are listed in roman font are multi-opposition, while the last
 five listed in italic font are single-opposition. In the latter case, the uncertainty
 of the proper elements is only formal because the uncertainty of the osculating
 elements may currently be larger.}
\end{table*}

The orbits of Hobson members were numerically integrated forward in time for
$2$~Myr using the base model, where only gravitational perturbation from
planets and the attraction by the Sun were included. In order to
enable a longer time step, we only discarded the innermost planet Mercury and
performed the corresponding barycentric correction in the state vectors of
planets. This has a negligible effect on the orbital zone of the main belt,
where the Hobson family is located. We implemented an online digital filter to
remove short-period terms (periods shorter than $300$~yr), so that the simulation
provided  the mean orbital elements for each of the bodies. The postprocessing
then represented the application of the Fourier analysis and the removal of forced (planetary)
terms, isolating thus the proper terms. Cartesian-like nonsingular elements
were used for $e_{\rm P}$ and $\sin I_{\rm P}$. To assess the formal uncertainty of
the proper elements, we used a simple running-box test. The nominal proper
elements were computed for the whole $2$~Myr long simulation. We then determined
these elements also on eleven $1$~Myr long windows shifted by $0.1$~Myr,
into which we segmented the original simulation. Their statistical standard 
deviation from the nominal values helped to characterize the formal uncertainty. 
We note that only nominal orbits of the asteroids were used in our simulation,
disregarding thus the current orbital uncertainty from the observations.
However, this is not a strong effect, except for the six single-opposition members
in the Hobson family: 2014~JH120, 2014~OJ66, 2020~JM31, 2020~KP36, 2020~OY50,
and 2021~MO5.
In the worst case, namely 2014~OJ66 with the poorest astrometric set, the realistic
semimajor axis uncertainty may be up to $\simeq 8\times 10^{-4}$~au.
Our results are summarized in Table~\ref{tab_prop} and Fig.~\ref{fa1}.

As in \citet{vetal2021}, we purposely used a short time-span of $2$~Myr in our
method to determine the proper elements. This presumably conforms to the very
young age of the family. However, there are drawbacks when longer-period terms perturb
the orbital evolution. The forced terms due to the Uranus or Neptune nodal or pericenter
precession are generally very small in the main belt. Some locations may be
affected by nonlinear secular resonances that involve the Jupiter and Saturn node
or perihelion precession frequencies, however \citep[e.g.,][]{mk1992,mk1994}. We find that
the zone of the Hobson family is significantly perturbed by the $g+g_5-2g_6$
secular resonance that produces $\simeq 1.2$~Myr oscillations in the eccentricity
vector and semimajor axis. The semimajor axis perturbation has a small amplitude, but
the eccentricity effect is strong enough to cause a high value of $\delta e_{\rm P}$
in our setup. Clearly, the values of $e_{\rm P}$ are perturbed themselves. In the same time,
their uncertainties $\delta e_{\rm P}$ are not statistically random, but again are systematically
affected by this resonance. Additionally,
the Hobson family is crossed by the three-body mean motion resonance 9J-8S-2 near
$a_{\rm P}\simeq 2.5636$~au \citep[see, e.g.,][]{gallardo2014}. This location is 
notable by (i) the observable uncertainty
of the $a_{\rm P}$ values and (ii) the anomalous dispersion of the $e_{\rm P}$ from our
$2$~Myr-long integration. This effect is not apparent to this degree in the
structure of the $\simeq 330$~kyr old Hobson family due to its youth, however. For these reasons,
the bottom panel in Fig.~\ref{fa1} is less useful for a discussion of the real Hobson family.
However, the top panel with its projection onto the $a_{\rm P}$ versus $\sin I_{\rm P}$ is
valuable and helps appreciate the compactness of the family well within the limits of
the $4$ m~s$^{-1}$ isotropic ejection field, supporting the low dispersal field of the
Hobson fragments studied in Sec.~\ref{conv}. The velocity limit in this projection
may easily be tightened to $\simeq 2$ m~s$^{-1}$, with one or two outliers explained by
an accumulated drift in proper semimajor axis by the Yarkovsky effect (up to
$\pm (2-3)\times 10^{-4}$ au in $\simeq 350$~kyr and subkilometer size body in the
Hobson family).

The impractical structure of the Hobson family in the $a_{\rm P}$ versus $e_{\rm P}$
adds to the consideration the usefulness of traditional proper elements in the case
of very young asteroid families. While a detailed analysis of this problem exceeds
the topic of this work, we note that the structure of the family in mean orbital
elements at the moment of convergence of secular angles (see Fig.~\ref{f4}) may
be a more practical tool. In Sec.~\ref{conv} we used information from mean inclination,
longitude of node, and perihelion. However, we also confirmed that near the
convergence epoch, the mean semimajor axis, eccentricity, and inclination may very
well replace the proper elements in a representation similar to that shown in Fig.~\ref{fa1}.
In particular, the position of the Hobson fragments again fit within the zone of
$4-5$ m~s$^{-1}$ isotropic ejection field. The long-period terms in these 
elements mean that the configuration of the family is not unique, however. Its compactness is
not surprising at this moment because we noted that the nodal and perihelion
dispersal values $\Delta\Omega$ and $\Delta\varpi$ contribution to the target function
(\ref{tf}) depend on all velocity components already.

\section{Setup of the SPH/N-body simulations}
The simulations presented in Sec.~\ref{sfd_mod} were performed with the following
numerical setup and material parameters (see also \citealt{sevecek2019} for more
details and definitions). The primary, the secondary, and the impactor were resolved
using $N = 10^5$, $10^4$, and $10^3$ particles. We used the \cite{Diehl_2012arXiv1211.0525D}
random-yet-isotropic initial distribution.

All materials were similar to monolithic basalt, with the density $\rho = 2700$ kg~m$^{-3}$,
the bulk modulus $B = 2.67\times 10^{10}$~Pa, the shear modulus $\mu = 2.27\times
10^{10}$~Pa, the elastic modulus $\epsilon = 8\times 10^9$~Pa, the
\cite{Tillotson_1962geat.rept.3216T}
equation of state parameters $a = 0.5$, $b = 1.5$, $B$ as above, $\alpha = 5$, $\beta = 5$,
an incipient vaporization energy $U_{\rm iv} = 4.72\times 10^6$~J,
a complete vaporization energy $U_{\rm cv} = 1.82\times 10^7$~J,
a sublimation energy $U_{\rm sub} = 4.87\times 10^8$~J,
an initial scalar damage $D = 0$, a von Mises rheology,
a von Mises limit $Y = 3.5\times 10^9$~Pa,
a melting energy $U_{\rm melt} = 3.4\times 10^6$~J,
a Weibull coefficient $k = 4\times 10^{35}$,
and a Weibull exponent $m = 9$.
We also performed tests with the Drucker-Prager rheology,
but if the pressure-dependent limit $Y(P)$ for a peak pressure~$P$
is similar to $Y$ above, the outcome is similar.
We did not analyse shapes of individual fragments.

The fragmentation phase duration was $10^3$~s.
The time step was controlled by
the Courant number $C = 0.2$,
the derivative factor $0.2$, and
the divergence factor $0.005$.
We used the asymmetric SPH solver,
the standard SPH discretisation,
the correction tensor for rotation,
the predictor-corrector integrator,
and we summed over undamaged particles.
The artificial viscosity parameters were $\alpha = 1.5$, $\beta = 3$.
We also used the Barnes-Hut gravity solver,
with the opening angle $\phi = 0.5$,
the multipole order $\ell = 3$,
and eventually, an equal-volume handoff.

The reaccumulation phase duration was $10^5\,{\rm s}$,
computed with the leap-frog integrator,
a ``merge-or-bounce'' collisional handler, and
a ``repel-or-merge'' overlap handler.
The derivative factor was $0.005$.
For the normal restitution, we assumed a value $0.5$,
a tangential restitution $1$,
and a merge velocity limit $\alpha_v = 0.25$,
where the condition for merging is
\begin{equation}
v_{\rm rel} < \alpha_v \sqrt{{2G(m_1+m_2)\over r_1+r_2}}\,;
\end{equation}
similarly, the merge rotation limit $\alpha_\omega = 1.0$.
The final SFD was computed from masses.
The model is still somewhat resolution dependent
because the number of particles determines the smallest block size,
and the SFD was built from these blocks.

\section{Contact binary model}
\begin{figure}
\centering
 \includegraphics[width=7.5cm]{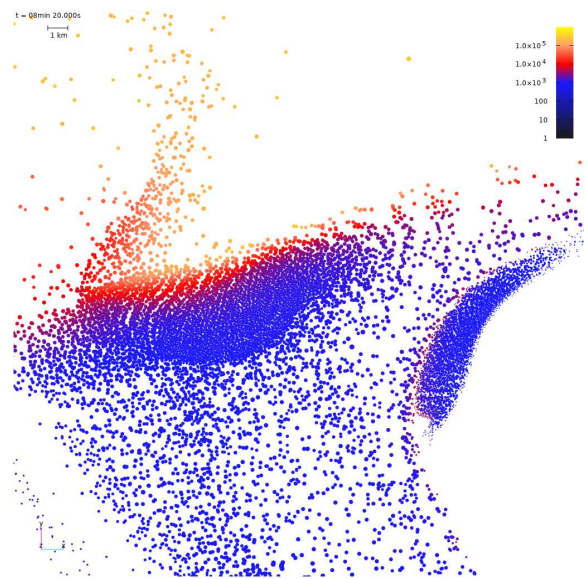}
\caption{\label{fc1}
 Simulation of a contact binary breakup. The spatial distribution of SPH
 particles is shown within a limited range of $z \in (-1, 1)\,{\rm km}$
 and for $t = 500$~s. Colors correspond to the specific internal energy $U$ in
 J~kg$^{-1}$. The primary was dispersed after a collision with the impactor (not
 shown). The secondary (on the right) was squeezed by a low-speed collision
 with the primary. In the reaccumulation phase, the primary and the secondary
 eventually provide the first and second largest remnants in the synthetic family.}
\end{figure}

For sake of completeness, we also considered an impact onto a contact
(or close) binary, as opposed to the well-separated binary system discussed in
Sec.~\ref{sfd_mod}. In particular, we used the same system and impact geometry
parameters as in Sec.~\ref{sfd_mod}, but assumed a separation of the primary
and secondary $r=(D_1+D_2)/2$, making them in contact. This is a different
regime (compared to the wide binary system) because the singular neck
in the system does not permit propagation of the impact-generated shock wave
into the secondary. Nevertheless, the latter was now affected to a much higher
degree. This is because the secondary was slowly but efficiently pushed by
the primary and squeezed along the
perpendicular direction (see Fig.~\ref{fc1}). All these motions were highly subsonic.
Eventually, most of the secondary mass was reaccreted because the mutual velocities
are relatively low compared to direct ejecta from the primary. The reaccreted
secondary must have a different internal structure, with damaged material and a fresh
surface, than in the model of wide binary parent system. This is different from the
case of a wide binary that we discussed in the main text.

\end{appendix}

\end{document}